# Software Requirements Engineering Healthcare Implementation Maturity Model (SRE-HIMM) for Global Health-Care Information System

Muhammad Hamza

**Abstract**

A Healthcare information system (HIS) is a complex information system in nature that deal with early disease detection and therapeutic. However, the development of these systems is quite complex in a collocated environment. Therefore, Healthcare organizations are outsourcing their projects to a globally distributed environment. Global software development has gained many intentions due to its economic and strategic effects on the development of the project. Though there are several benefits of outsourcing projects GHIS also poses several challenges particularly pertaining to the HIS development process.

Several Process improvement techniques and standards have been developed to assist organizations in their effective management of the development process. For example, Capability Maturity Model Integration (CMMI) is widely accepted by organizations for their process assessment and improvement. Similarly, the International Organization for Standardization (ISO) i.e., ISO 9000 used for the assessment of the quality of the developed system.

The deployment of HIS process improvement in the globally distributed environment requires adequate resources and time. Though working in a distributed environment makes communication and collaboration difficult and developing the HIS process improvement program is much pronounced. HIS teams face difficulties establishing relationship among other members, eliminating the temporal and cultural boundaries become significant.

The fundamental objective of this research work is to develop a Software Requirement Engineering Healthcare Implementation Maturity Model (SRE-HIMM) that will assist healthcare organizations to effectively evaluate and implement HIS development process. The model was developed based on the systematic literature review (SLR) approach and Empirical results. The 53 primary studies were extracted using the SLR approach and CSFs, CBs, and best practices were identified form the extracted primary studies. The identified success factors and barriers were further ranked using the analytical hierarchy process (AHP) approach. Furthermore, I have adopted the critical success factors (CSFs) and critical barriers (CBs) instead of PAs and available Maturity models i.e., CMMI for the development of (SRE-HIMM). The identified CSFs and CBs were classified into five maturity levels based on the CMMI, IMM, and SOVRM. The empirical investigation was conducted HIS experts to evaluate the findings of SLR. Further, a case study was conducted with the company to evaluate the effectiveness of SRE-HIMM which shows satisfactory results.

# Chapter 1: Introduction

### 1.1 Requirement Engineering (RE)

Software-intensive systems have penetrated nearly all aspects of our lives, in a huge variety of ways. Information technology has become so powerful and so adaptable, that the opportunities for new uses seem boundless. However, our experience of actual computer systems, once they have been developed, is often disappointing. They fail to work in the way we expect, they are unreliable, and sometimes dangerous, and they may create more problems than they solve. The success of any software system depends on how well it fits the needs of its users and its environment. Software requirements comprise these needs, and requirements engineering (RE) is the process by which the requirements are determined. A requirement is a property that a system must exhibit to meet the system's motivation need, and software requirements are a property that must be exhibited by software developed to solve a particular problem within one organizational context. Therefore, the software requirements are a complex combination of requirements from different people at different levels of an organization and from the environment in which the system must execute. They express the needs and constraints placed on a software product that contribute to the solution of some real-world problems and normally result in an arrangement between "user requirements" and "system requirements". User requirements denote the requisites of the people who will be the system client or end-users. System requirements add requirements of other stakeholders (such as regulatory authorities) and requirements that do not have an identifiable human source and that normally result from the intersection among technical, cultural, and social environments (Lam, W. Shankararaman, V., 1999; Lopez, A. Nicolas, J. Toval, A., 2009, Sardar et al 2022, Hamza et al 2022). Successful RE involves understanding the needs of users, clients, and other stakeholders, as well as understanding the context in which the software will be used.

In general, the RE is the process of determining the goal, by investigating the expectations of stakeholders and documenting these expectations in such a way that can be analyzed and implemented. Brook has identified the important and complicated phase of software engineering for the successful development of systems or software. The main purpose of these phases is to determine exactly what to build, who will be involved in the system, how actual functionality will be performed. Most of the systems fail due to the improper management and implementation of the phases of software developments and one of them is requirements gathering. The software development life cycle (SDLC) consists of different phases i.e. Preliminary Investigation, Analysis, Design,

Development, Testing, Implementation, maintenance (Akbar et al 2022, Khan et al 2022, Rafi et al 2022, Riaz et al 2022, Qadri et al 2022). The first and foremost important phase is requirements gathering for the successful implementation of software or a system. Improper identification and management of requirements for the system could destroy fully or may cause an unbearable financial loss. Similarly, (Khan et al. 2014, Akbar et al 2022) has investigated the first phase of system development life cycle (SDLC), the requirement engineering and decomposed the RE process into five core phases (i.e. "requirements extraction or elicitation, requirements analysis and design, requirements specification, requirements validation, and requirements management"). The requirements could be changed when the development of the system begins. The customer could demand a new change in the system when it is developed. Thus, change in requirements at a later stage could cause financial loss, and change in the requirements at a later stage could be challenging to manage efficiently and effectively.

Requirements change management is defined as the management of changing requirements during the requirements engineering, system development, and the maintenance process (S. Jayatilleke and R. Lai 2018, S. Ramzan and N. Ikram 2005). The change could be difficult or could be expensive while implementation at a later stage. For example, (Nurmuliani et al 2014) have stated that the change in requirements may occur due to the increase in stakeholder understanding and change in the organizational work environment (Akbar et al 2022).

Research has shown that improper management of requirements changes while developing the large-scale system can lead to project failures (S. Ramzan and N. Ikram 2005) and result in software cost blowout, schedule overrun, and potential business loss. Several empirical studies have been carried out to discuss the management of requirements change in a traditional in-house software development context. For example, (Lai et al. 2011 and Akbar et al 2022) underlined that for developing the quality product the pure requirement collection and management process is significant. According to (Bano et al. 2010, Mehmood et al 2021, Akbar et al 2022) the requirements change occur due to the changes in market demand, customer need, change in government, or organizational policies, due to software complexities and due to increasing the understanding of clients and practitioners. (Khan et al 2014, Rafi et al. 2022, Akbar et al. 2019, Akbar et al. 2020, Akbar et al 2022) have investigated that requirements collection and management demand rich communication among the client and development team. Due to demanding the rich communication and collaboration in the requirements process, RCM is considered as challenging in the context of collocated (single site) software development environment. Therefore, Software organizations are distributing their work around the globe to achieve skilled labor and to manage their resources. Several the software organizations converting their businesses from collocated to geographically distributed development environment due to the diverse nature. A survey study conducted by Standish Group indicated that 20% of the client software organization outsourcing their development activities to vendor countries (C. Manifesto 2014). (Smite et al. 2010, Akbar et al 2022) argues that the GSD environment assists the software firms to develop quality software at low cost and time. They further highlighted that the GSD provides the opportunity of time to market and the availability of the latest tools and technologies.

(Assawamekin 2010) conducted a survey and highlighted that during software development 44% to 80% of bugs are occurred due to poor requirements management mechanism. The poor requirements management causes the budget overrun and time overrun and most of the time it leads towards the project failure (S. Ramzan and N. Ikram 2005). (Lindquist 2006) presents an analyst report and highlighted that 71% of software projects are failing due to the poor management of demanded requirements changes. The lack of effective requirement management is one of the biggest problems of successful software projects. Similarly, (Sirvio and Tihinen 2005) surveyed with European software development organization and underlined that 40% of software projects failed due to poor requirements management.

## 1.2 Healthcare Information System (HIS)

The use of sophisticated information and communication technologies (ICTs) in the health care domain is a way to improve the quality of services. The literature points to a consensus that health information systems (HISs) are thought to have the potential to improve patient care. However, there are also hazards associated with the introduction of ICTs in health care, and some works report how difficult it is for the successful introduction of ICTs in this domain. Health care is a unique and complex domain and HISs have human safety implications and profound effects on individual patient care. The successful development of HISs can increase efficiency and productivity, ease of use and learning, adoption, retention, and satisfaction of the users, and simultaneously, it can help to decrease medical errors as well as to reduce support and training cost. On the other hand, HISs are usually complex systems and their failures may cause negative effects on patients, and possibly, when insufficiently designed, they may result in spending more time with the computer than with the patient. According to (C. Manifesto 2014) ICTs have been hailed as a solution to reduce errors in health care, but there is also evidence that they can be part of the problem. The system collects data from the health sector and other relevant sectors, analyses the data and ensures their overall quality, relevance, timeliness, and converts data into information for health-related decision-making" (Ziefle, Martina, and Anne Kathrin Schaar 2011). The goal of using health information systems is to improve patient treatment by having the most current patient data available to every healthcare practitioner who treats this client. Therefore, health-care information system has gained popularity tremendously as it improved the clinical decision making and planning process, the health-care management system has been developed and deployed to several areas for patient's monitoring and management. Health information systems are available too and accessed by, healthcare professionals. These include those who deal directly with patients, clinicians, and public health officials. Healthcare professionals collect data and compile it for use in making health care decisions for individual clients, client groups, and the public. Healthcare information systems include:

**Electronic Medical Record (EMR) and Electronic Health Record (EHR)** "these systems records, replace paper patient records. The medical information on each patient must now be collected and stored electronically. These records would include patient health information, test results, doctor and specialist visits, healthcare treatments"

**Practice Management Software** "assists healthcare facilities and personnel with the management of daily operations of the facility. This would include things like scheduling of patients and medical services billing"

**Remote Patient Monitoring (RPM)** "provides medical sensors that can transmit patient data to healthcare professionals who might very well be halfway around the world. RPM can monitor blood glucose levels and blood pressure. It is particularly helpful for patients with chronic conditions such as type 2 diabetes, hypertension, or cardiac disease"

**Clinical Decision Support (CDS)** "analyzes data from clinical and administrative systems. The aim is to assist healthcare providers in making informed clinical decisions. Data available can provide information to medical professions who are preparing diagnoses or predicting medical conditions like drug interactions and reactions".

**Master Patient Ind0ex (MPI)** healthcare information system is aimed at connecting patient records more than one database. The MPI contains records for any patient registered at a healthcare organization. MPI, as the name suggests, creates an index all the records for that patient.

Although, while developing these kinds of systems there needs to involve the requirement engineering phase of software or system development life cycle (SDLC), as it is considered the foundational part of every system before the initiation of developing and during the development process.

## 1.3 Global Software Development for Healthcare Information System (GSD)

The development of Information systems and software is getting more and more globally distributed. The economic factor remains the most influential driver of this phenomenon. The continuous growth, innovation, and ongoing improvements in ICT's enable even complex projects and systems to get developed at places with geographical, temporal, and cultural distance. Cost reduction, increased production, risk dilution, and improvement in quality as well as flexibility in software development are the means to competitive advantage and are the motives common to the software industry worldwide. The inclination towards Global Software Development (GSD) is obviously because of its well-identified and documented benefits (Conchuir et al. 2009) that include cost savings, access to large multi-skilled workforces, proximity, and reduced time to customer market, etc. (Conchuir et al. 2009). The quest for business excellence and competitive advantage compels organizations to look for solutions around the globe (offshore sourcing or offshoring). GSD appears as a feasible alternative (Prikladnicki et al. 2006) in such an environment. However, during that time large investments paved the way towards the movement of globalization which then resulted in the creation of new forms of competition and collaborations (Prikladnicki et al. 2006, Akbar et al 2020). The idea of globally distributed software development, therefore, continues to gain momentum. (Sahay 2003) defines GSD as follows:

*"Global software development is the software work undertaken at geographically separated locations across natural boundaries in a coordinated fashion involving real-time (asynchronous) and synchronous interaction"*

GSD is facing more problems with changing requirements and their management as compared to single-site development. Despite the significant benefits associated with GSD, it also faces a set of unique challenges that do not exist while developing the software in a single-site development context. (Khan et al. 2014, Akbar et al 2020) underlined that language and cultural differences, lack of face to face meetings, time-zone differences, and delay in overseas sites response are considered as critical barriers in the RCM process in GSD. Similarly, (Niazi et al. 2016, Akbar et al 2021) state that physical separation between GSD teams and lack of frequent information sharing hinders the successful implementation of the RCM process in GSD.

Several studies have been conducted to address the complexities of RCM activities in the domain of global software development (GSD). (McGee and Greer 2009) and (Pierce et al. 2013) highlighted that the demanded changes in requirements are not a problem, and the problem is that how effectively address the changes. (Nurmuliani et al.2006) emphasized that the key step towards the management of demanded changes in the identification of the root cause of demanded change. (Barry et al. 2002) highlighted that the uncertainties in the initially collected requirements and the lack of customer's involvement causes the change in requirements. (Chrissis et al.2006) argued that the history of changes in requirements should be recorded. They further suggest that the requirements change history should be documented and considered to identify the volatility in requirements. (Ngwenyama and Nielsen 2008) emphasized that in GSD the physical distance between the practitioners causes the lack of face to face communication and the cultural distance causing misunderstandings between the overseas practitioners. Thus, the problems of RCM in GSD are more complex than the collocated development environment (A. A. Khan 2014, Akbar et al 2021). (Esbensen et al. 2014) emphasized the lack of face to face meetings with GSD practitioners causes a lack of trust. Similarly, a survey study conducted by (Ramzan and Ikram 2006) argued that the RCM process area is not standardized even in the collocated software development environment, this renders the lack of RCM process in the context of GSD.

This study is carried out owing to the need for investigation of RCM processes as pointed out by various other researchers. For example, (Lam et al 1999), suggests that in the software industry the collective guidance for managing requirements change is still weak and there is a need for developing "systematic and methodical practices for managing requirements change". Many partial solutions have been offered for the implementation of RE in a GSD environment, but they lack process-level detail (Lopez et al. 2009). There remains a gap in this area to be filled up with more rigorous research on the RCM process. Without a Requirements Engineering (RE) process suitable for GSD, specially designed for Requirements Change Management (RCM) it is difficult to avoid the challenges global software development is faced with (Sangwan et al 2007, Akbar et al 2021). Other studies also hint towards having a whole new set of techniques and strategies that are required to successfully carry out GSD projects. Similarly, (Zowghi 2007) suggests that elicitation, analysis, specification, validation, and management of requirements remain one of the least explored and have the least satisfactory scientific foundations. Considering the novelty of the GSD paradigm and the lack of research work in the RCM process in the GSD environment.

## 1.4 Research Objective

The ultimate objective of this study is to develop a maturity model that will assist organizations to assess and evaluate their development process especially in the domain of healthcare information system (HIS) by investigating the literature and expert's opinion.
The model is developed based on the HIS development process improvement literature, industrial empirical study, and factor that could positively and negatively affect the development process.

## 1.5 Research Questions

Six questions motivated us to work on.
RQ1. What are the factors reported in the primary literature, do GHIS organizations need to address reported factors for the successful implementation of HIS process?
RQ2. What are the barriers reported in the literature, does it really have a negative effect on HIS process Implementation?
RQ3. What are the main success factors reported in the real practice, do GHIS organizations need to address for the successful implementation of HIS?
RQ4. What are the best practices that are reported in the primary studies to address the critical factors and critical barriers?
RQ5. What are the barriers, as identified in the real practice, within GHIS have a negative impact on HIS implementation?
RQ6. What are the practices, as identified in real practice, to address the critical success factors and critical barriers for HIS in GHIS?

## 1.6 Thesis Structure

The thesis is organized as follows.
Chapter 2: This chapter focuses on the previously done studies, that address the requirement engineering perceptive for the healthcare information system (HIS).

Chapter 3: Presents the research methodology towards the development of the Software Requirement Engineering Healthcare Implementation Maturity Model (SRE-HIMM). This chapter briefly discussed the selected research methods and their justification. It also contains the adopted statistical techniques selected to analyze the collected data.

Chapter 4: This chapter reports the results of the SLR, and the empirical study conducted with the HIS experts. This chapter identified the CSFs and CBs along with their practices.

Chapter 5: Highlight the complete development process of SRE-HIMM by describing the several components.

Chapter 6: Represent the summary of all research questions.

# Chapter 2: Literature Review

This chapter discusses the basic terms and definitions used in the health care information system and Global Software Development. It introduces the role of Requirements engineering in the health care information system and the concept of Globalization and Global Software Development and talks about the benefits it provides as well as the challenges it faces. Then discuss the existing requirements engineering literature in the domain of healthcare information system.

## 2.1 Definitions of Relevant Terms

For understanding the reader, we define the most commonly used terms used in the health care information system and requirements change management process when performed in a globally distributed environment.

**Software Requirement** as given by Dorfman and Thayer (1990) cited by Leffingwell and Widrig (2003) is "a software capability needed by the user to solve a problem to achieve an objective". Alternatively; "a software capability that must be met or possessed by a system component to satisfy a contract, standard, specification, or other formally imposed documentation". Finally, we present a compact and workable definition by Oberg (2000) that "a requirement is a condition or capacity that a system that is being developed must satisfy".

**Requirements Change (RC)** refers to "the emergence of new requirements or the modification or removal of existing requirements" (Lam & Shankararaman 1999). Changing requirements are considered to be a cause of failure for new as well as legacy systems as both have to go through several requirement changes (Lam et al 1999). Requirements change encompasses:

**Requirements volatility** which is a term meaning 'a measure of the number of requirements changes (addition, deletion, and modification) (Davis 2005 Cited by Davis et al., 2008) divided by the number of requirements for a given period (Costello et al 1995 cited by Davis et al 2008).

**Requirements Creep** which refers to 'frequent changes in requirements' (Jones 1996 Cited by Davis et al 2008), and also described as changes that result "in extensions to and alterations of the software's functionality and scope' (Carter et al 2001 cited by Davis et al 2008)

**Requirements Management (RM)** means "the systematic process of organizing and storing relevant information about requirements while ensuring requirements traceability and managing changes to these requirements during the whole lifecycle of the information system" (Grehag 2001).

**Requirements Change Management (RCM)** is concerned with making rational decisions about whether to implement a requested change or not. It is also concerned with supporting the identification of which information, e.g. documents and other requirements that are affected by the proposed change (Grehag 2001). Change management is not easy to perform even under the best of circumstances (Sangwan et al 2007) and it becomes more difficult when performed globally because of the nature of distributed development projects and the diversity of stakeholders (Damian & Zowghi, 2003).

## 2.2 Requirement Engineering in Health-care Information System.

The successful development of HISs can increase efficiency and productivity, ease of use and learning, adoption, retention, and satisfaction of the users, and simultaneously, it can help to decrease medical errors as well as to reduce support and training cost. On the other hand, HISs are usually complex systems, and their failures may cause negative effects on patients, and possibly, when insufficiently designed, they may result in spending more time with the computer than with the patient. According to (J. Aarts et al 2007) ICTs have been hailed as a solution to reduce errors in health care, but there is also evidence that they can be part of the problem. There are a large number of HISs projects that have failed, and most of these failures are not due to flawed technology, but rather due to the lack of systematic consideration of human and other non-technology issues throughout the design or implementation process. Also, several studies have shown that 80% of total maintenance costs with information systems (ISs) are related to users' problems and not technical bugs, and among them 64% are related to usability problems (B.W. Boehm 1991). A survey of over 8000 projects made by The Standish Group and undertaken by 350 US different companies revealed that one-third of the projects were never completed, and one half succeeded with partial functionalities. The major source of such failures resided on poor requirements, specifically: the lack of user involvement (13%), incomplete requirements (12%), changing requirements (11%), unrealistic expectations (6%), and unclear objectives (5%) (Standish Group 1995). These problems are mainly because, in developing interactive software, most software engineering methodologies do not propose any mechanisms to: (i) explicitly and empirically identify and specify user needs and usability requirements; and (ii) test and validate requirements with end-users before and during the development process (A. Seffah 2005). The health care domain has been particularly prone to such problems in recent years, and there are numerous examples of potentially useful systems that have failed or have been abandoned due to unanticipated human or organizational issues. Since the design of systems that are used by people is a complex endeavor, the systems that cannot be used intuitively often lead to an increase in error rate and a decrease in user acceptance. According to (Leonor Teixeira 2012) the principles of user-centered design (UCD) combined with ethnographic practices can improve synergies among technology, people, and work environment (tasks). Therefore, to perform a user needs analysis and to write requirements specification for integrated care in the hemophilia field, they have followed a user-centered requirement engineering process involving the end-users through different techniques of requirements elicitation and validation. Similarly, (R. Lenz et al. 2004) have addressed the problem of alignment of health information systems to healthcare processes, which is a major challenge in healthcare organizations and have presented a layered approach for system evolution and adaptation based on an application framework and rapid application development and accomplished a demand-driven system evolution by embedding the software engineering process in business process optimization projects and by closely involving end-users to improve their work practices. Similarly, Grace (A. Lewis et al.2009) described that traditional requirement engineering for single systems, while remaining a large challenge for engineers, has been extensively researched and many techniques have been proposed and used with varying degrees of success. However, many modern systems of systems (SoS) are being developed to support interaction across multiple controlling authorities and existing techniques are proving to be inadequate for meeting the challenges of requirements engineering for systems of systems in the healthcare domains. (Hongqiao YANG et al. 2010) described that to improve the quality of healthcare services, large-scale medical information systems should be integrated with adaptability in response to the changing medical environment. To address the issue, they have proposed a requirement-driven architecture for healthcare information systems that will be able to respond to new requirements. The system operates through the mapping mechanism between these layers and thus can organize functions dynamically adapting to the user's requirement. (George M. Samaras et al. 2005) described that the classical SE method, emphasizing that it provides a framework for incorporating ergonomics knowledge in all phases of the interdisciplinary development process and integrating the role of ergonomists into the development team. They have compared and contrasted the classical systems engineering method to more recently published development lifecycle methods, pointing out that the latter represent incomplete subsets of the former. Similarly, (I. Scandurra et al. 2008) introduced a new multi-disciplinary method for user needs analysis and requirements specification in the context of health information systems based on established theories from the fields of participatory design and computer supported cooperative work (CSCW). (Rita Noumeir 1970) described that Interoperability is a requirement for the successful deployment of Electronic Health Records (EHR). EHR improves the quality of healthcare by enabling access to all relevant information at the diagnostic decision moment, regardless of location. It is a system that results from the cooperation of several heterogeneous distributed subsystems that need to successfully exchange information relative to a specific healthcare process. (Marilyn Rose et al. 2008) discussed the features that characterize home care and illustrate the complexity of the home care domain. It, therefore, argues that a novel RE approach is required for the development of home care technology and suggests several features that should be available in requirements engineering for home care technology. (Xiping Song et al. 2006) as the requirement engineering and prototyping group of the Siemens R&D center, have been involved in the research and development of healthcare workflows. During their interactions with the workflow users and developers, they found significant confusion about the terminologies and the purposes of supporting different healthcare workflows. (Ian Sommerville 2003) discussed an approach to system requirements elicitation that integrates safety requirements elicitation and analysis with more general requirements analysis. he has proposed that the analysis should be organized around pervasive 'concerns' such as safety and security which can drive the requirements engineering process. (Toomas Timpka et al. 2008) have argued and conducted a case study that the development of computer-based information systems in public health is thus more than simply a technical challenge; it requires consideration of behavioral, social, and organizational issues associated with information management. They have examined the need for information systems support in modern health promotion programs, using the international Safe Community program as an example. Similarly, (Sebastian Garde et al. 2006) discussed that health care is characterized by highly complex processes of patient care that require an unusual amount of communication between different health care professionals of different institutions. they have presented and discussed an approach to requirements engineering that they applied for the development of applications for chemotherapy planning in pediatric oncology. (L. Teixeira et al. 2007) proposed a method for eliciting requirements based on participation, collaboration, and negotiation of requirements by the different stakeholders of the project, supported by the method of prototyping. This method promotes an approach of stepwise refinement of requirements, starting with general ideas until the low-level requirements are achieved. Since this is a prototype, this method was also essential in the identification of some non-functional requirements that complemented the definition of the functional requirements.

# Chapter 3: Research Methodology

## 3.1 Overview

This chapter provides a brief description of the various research philosophies, particularly qualitative and quantitative research methodology. A systematic literature review (SLR) methodology is also explored and is used to gather data from the literature. This chapter consists of the adopted research techniques and their justification. The data collection and analysis processes will also be explained in detail.

This chapter is organized as follows:

• Section 3.2 gives a philosophical view of the research methodology.

• In Section 3.3, the background of the research methodology is discussed.

• In Section 3.4, choosing a research method and justification is reported

• Section 3.5 consists of data collection approaches.

• In Section 3.6, data analysis techniques are discussed.

• Section 3.7 consists of limitations of the selected research methodologies.

• In section 3.8, the chapter is concluded

## 3.2 Philosophical View

The most critical prerequisite for all research studies is the correct collection of methods to be applied to research questions (Kothari, Hu, & Roehl, 2007). The most relevant methods are those related to the underlying epistemology that drives the research process. According to (Audi, 2010), epistemology is an assumption about knowledge and how it can be acquired. Two categories that are based on the underlying research epistemology are Positivist and Interpretivist.

### 3.2.1 Positivist

According to Levin (1988), a positivist approach to study is focused on information obtained from 'true' evaluation of measurable experience, e.g. introspection or intuition. Scientific approaches or experimental experiments are the best way to gain this understanding. Easterbrook et al. ( 2008) argue that positivist views of the world in order and on a regular basis can investigate its objective. Positivists also believe that the observer is not a part of research and science is not driven by human interest.

### 3.2.2 Interpretivist

Interpretive methods of study are focused on the assumption that our understanding of reality, including the field of human action, is a collective creation of human actors and extends equally to researchers. Thus, there is no objective reality that can be discovered by researchers and replicated by others, as opposed to the assumptions of positivist science (Walsham, 2006). Audi's report (2010), The interpretive approach is more focused on identifying, exploring, and explaining the full complicity of human sense-making as the situation emerges rather than a defined hypothesis. They added that the goal of interpretive researchers is plausibility rather than evidence, as in the case of positivists. The Positivist paradigm is more about the generalizability of previous work than an interpretive model that could enhance the depth of understanding of the phenomenon in question (Walsham, 2006).

## 3.3 Research Methodology background

Research methods are the means of collecting data, such as surveys and interviews, while research methodologies refer to strategies used to conduct research activities to generate reliable research outputs (Bryman, 2012). Qualitative and quantitative are two concepts used to distinguish the nature of analysis (Creswell, & Clark, 2010). Qualitative analysis methods are based on an interpretivist paradigm, while quantitative methods are based on the positivism paradigm and (Firestone, 1987).

### 3.3.1 Quantitative research

The Quantity Analysis Method is a methodology that describes the selected phenomenon by gathering, evaluating, and recording the effects of numerical records using different statistical methods (Creswell, 1994). Most researchers have adopted a quantitative study technique to conclude whether or not the generalization of a theory is true or not (Creswell, & Clark, 2010). It may help to determine the relationship between dependent and independent variables by proposing and testing study hypotheses (Bryman, 2012). Various forms of quantitative research (Sukamolson, 2011) are discussed briefly in the following sections.

#### 3.3.1.1 Survey research

The definitions of the survey and survey analysis have a significant distinction. A survey is a way of "gathering information on the attributes, behavior, or views of a broad number of people referred to as a community" (Adams, 1982). As such, various data collection methods are called surveys, e.g. research surveys, marketing surveys, and opinion surveys. Research conducted by the survey focuses on the advancement of scientific knowledge (Rea, & Parker, 2014). Survey research produces objective data using many people who need to be a sample to test the hypothesis.

#### 3.3.1.2 Correlation Research

Correlation research focuses on the study of the relationship between independent and dependent variables (Borrego, Douglas, & Amelink, 2009). Correlation analysis is the extent to which the variables are interlinked. The correlation coefficient (r-value) is used to measure quantitatively the extent of the relationship (Stangor, 2010).

### 3.3.1.3 Experimental Research

Experimental studies technique is ordinarily followed in natural sciences studies along with medicine, physics, etc. It is a group of research designs that use manipulation and managed to try out to understand the causal processes (Bryman, 2012). In experimental studies, the researchers manipulate one or greater variables to determine their effect, control, and change on a structured variable (Wiersma, & Jurs, 2005).

### 3.3.1.4 Causal-comparative Research

Causal comparative research is just like descriptive studies because it describes the situations that already exist (Sukamolson, 2011). In causal-comparative research, the researcher tries to decide the reason for the present differences among the behavior of groups or individuals. Established corporations which might be already exclusive for a few variables and the researcher's tries to pick out the major factor that has caused this difference (Abawi, 2008).

### 3.3.2 Qualitative research

Most of the researchers, working in the social sciences, were interested in analyzing human conduct and the social international occupied by human beings (Glaser, & Strauss, 1967). They determined increasing issue in looking to give an explanation for human conduct in simple measurable terms (quantitative studies methods) (Hennink, Hutter, & Bailey, 2010). Measurements tell us how often or what number of people behave in a certain manner, but they do no longer sufficiently solution the question "why?" (Bryman, 2012). Research which tries to boom our information of why matters are the manner they're in our social global and why human beings act the ways they do is "qualitative" studies (Bryman, 2012). Qualitative studies are worried about locating the answers to the questions which start with: Why? How? In what manner? Quantitative studies, on the other hand, are greater concerned with questions about: how much? How many? How regularly? To what extent? (Sekaran, & Bougie, 2010). The 4 major sorts of qualitative studies layout are categorized as follows (Silverman, 2010):

### 3.3.2.1 Phenomenology

The word phenomenology has been used to research phenomena. It explains the things that already exist in this universe. Phenomena can be situations, circumstances, interactions, or expectations (Silverman, 2010). There are a lot of phenomena in this universe, but we do not have sufficient awareness and understanding of these phenomena. Phenomenology research does not need to provide reliable knowledge, but it simply raises consciousness and increases insight.

### 3.3.2.2 Ethnography

Ethnography has an anthropology background. The phrase "portrait of a human" is a tool for the descriptive study of cultures and peoples (Bernard, 2011). The cultural parameter is that there is something in common for the people under investigation. Types of criteria include geographical-specific regions or nations, political, cultural, mutual interactions, etc. (Bernard, 2011). Ethnographic study may be problematic if researchers are not fully acquainted with the social practices of the people being studied or with their language.

### 3.3.2.3 Grounded theory

Grounded Theory method (GT) is a systematic research method that used to find out the concept by studying the focused data (Glaser, & Strauss, 1967). It goes past phenomenology due to the fact the explanations that emerged are virtually new knowledge and are used to expand new theories approximately a phenomenon (Carver, 2007). Case study research is used to explain an entity that bureaucracy a single unit consisting of a person, an organization, or an institution. Some research describes a chain of cases (Glaser, & Strauss, 1967).

### 3.3.2.4 Action Research

Action research is a strategy educator can use to observe educational issues, implement change, and record professional growth (Munn-Giddings, & Winter, 2013). Action studies are either research initiated to clear up an immediate hassle or a reflective process of progressive problem fixing led via individuals working with others in teams or as a part of a "network of practice" to improve the manner they address troubles and resolve problems (Munn-Giddings, & Winter, 2013).

### 3.4 Selected research methods and justification

In empirical studies, I even have used each qualitative and quantitative approach (mixed method) to gather the records to create and take a look at the variety of studies questions. In the mixed studies technique, both the quantitative and qualitative statistics had been amassed concurrently in an unmarried study (Borrego, Douglas, & Amelink, 2009). The mixed research approach assists to crush the constraints of each qualitative and quantitative research approach (Creswell, & Clark, 2010). The implementation of quantitative and qualitative techniques involved statistics collection that divided into two approaches: sequential or concurrent with the priority to one technique over the opposite of having the identical and same status (Creswell, & Clark, 2010). Qualitative and quantitative facts series techniques are complementary (Walker et al., 2003; Khan, 2011). Qualitative facts can be transformed into quantitative via the use of the coding method and this procedure will no longer affect the subjectivity or objectivity of the statistics (Baddoo, & Hall, 2002; Burnard, 1991; Khan, 2011). Bryman (2012) referred to that the reverse can also occur. One of the examples wherein quantitative research can facilitate qualitative research is through the selection of case research for further studies. The brief assessment of both quantitative and qualitative research techniques specifies that empirical research techniques can help researchers in well-founded decisions (Perry, Porter, & Votta, 2000).

In this research, I actually have used the mixed technique method for data series and the same technique has been followed through different researchers in exceptional different domains (Niazi, 2004; Khan, 2011; Khan, 2015). The qualitative facts were gathered using SLR and survey questionnaire approaches. The amassed qualitative have been transformed into quantitative (i.E. Frequencies) to carry out the statistical analysis. The overall studies system is depicted in Figure 3.1

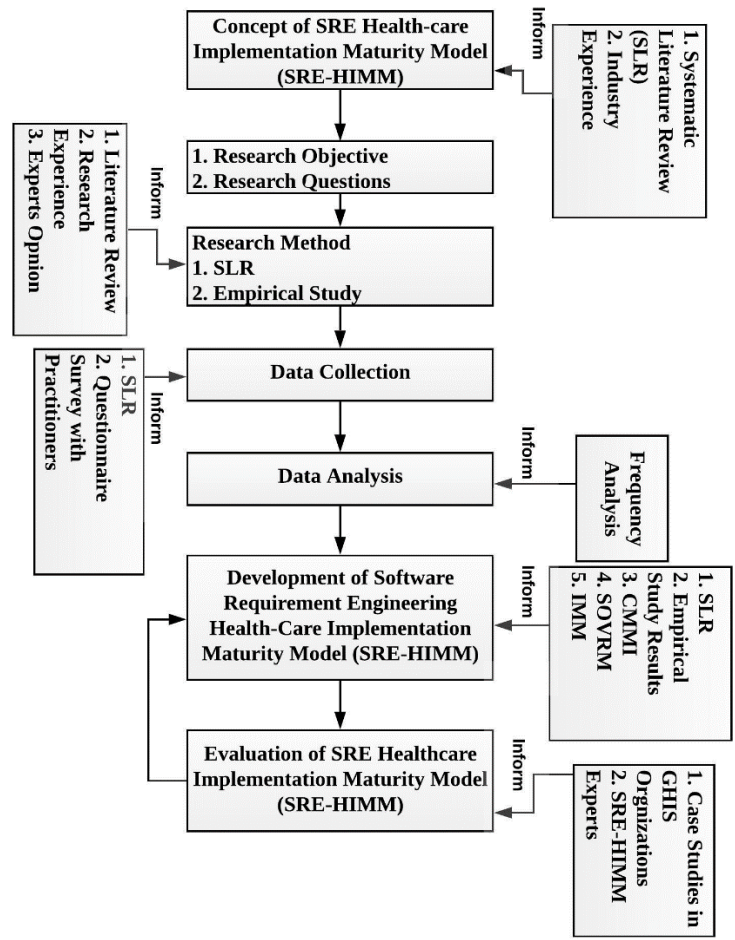

Figure 3.1 Research Process

## 3.5 Data collection

The data collection methods must be without a doubt defined and justified because it has a tremendous impact on the analysis process (Niazi, 2004; Khan, 2011, Akbar et al 2021). I have used an SLR and a questionnaire survey to collect records from existing literature and practitioners. These approaches were chosen because they are great for the type of facts analyzed and reported in this study (Rockart, 1979; Niazi, 2004; Khan, 2011, Akbar et al 2021, Yasir et al 2020, Kamal et al 2020, Muhammad et al 2020, Nawaz et al 2020). Systematic literature evaluation (SLR) is the systematic way of reporting the outcomes extracted from literature (Kitchenham, & Charters, 2007, Akbar et al 2021). SLR technique provides a way of exploring, classifying, and studying the current studies related to any studies place and questions of interest (Kitchenham, & Charters, 2007, Akbar et al 2021, Hayat et al 2021, Fahad et al 2021, Rafi et al 2021, Khan et al 2021, Mehmood et al 2021). The predominant goal of SLR is to seek out maximum relevant literature through applying specific inclusion and exclusion criteria at primary research and also to outline the high-quality standards for evaluation of the number one results and extracted facts from those studies (Kitchenham, & Charters, 2007). In this thesis, the guidelines provided by way of Kitchenham and Charters (2007) had been accompanied to carry out SLR. According to Kitchenham and Charters (2007), SLR consists of 3 levels i.e. making plans the evaluation, conducting the evaluation, and reporting the evaluation as shown in Table 3.1.

Table 3.1 SLR phases

| Planning the Review/Review Protocol development | <ul><li>Research questions</li><li>Search strategy and search terms</li><li>Inclusion and Exclusion criteria</li><li>Data sources</li><li>Quality Criteria for studies selection</li></ul> |
|---|---|
| Conducting the Review | <ul><li>Primary Studies Selection</li><li>Data Extraction</li><li>Data Synthesis</li></ul> |
| Reporting the Review | <ul><li>Document all the extracted results</li></ul> |

### 3.5.1 Planning the review/Review protocol development

Planning the review consists of different other sub-phases to develop the SLR protocol.

#### 3.5.1.1 Research Questions

I used the SLR approach to answer the following research questions:

RQ1. What are the factors reported in the primary literature, do GHIS organizations need to address reported factors for the successful implementation of HIS process?

RQ2. What are the barriers reported in the literature, does it really have negative effect on HIS process Implementation?

RQ3. What are the best practices that are reported in the primary studies to address the critical factors and critical barriers?

### 3.5.1.2 Search strategy and search terms

The following approach has been evolved to assemble search terms (Khan et al., 2013):

a. Derive the principal phrases from the research questions, by figuring out population, intervention, and outcome of relevance.
B. Derive the fundamental terms and search for their alternative spellings and synonyms.
C. Verify the primary terms in any applicable papers.
D. If the database allows, use the Boolean "OR" operator to integrate the alternative spellings and synonyms and use the "AND" operator to combine the main terms. I even have assessed each study question consistent with the above search approach.

(RQ1) What are the factors reported in the primary literature, do GHIS organizations need to address reported factors for the successful implementation of HIS development process?

**Search String for success factors:**

The keywords and their alternatives were based on existing literature in the domain of software process improvement and GSD (Babar, & Niazi, 2008; Ramasubbu, 2014; Khan et al., 2011; Khan, & Keung, 2016).

Factors: ("factors" OR "aspects" OR "items" OR "elements" OR "drivers" OR "motivators" OR "variables" OR "characteristics" OR "parameters") AND ("GSD" OR "global software development" OR "global software engineering" OR "distributed software development" OR "software outsourcing" OR "offshore software development" OR "information technology outsourcing" OR "software contracting-out" OR "IT outsourcing" OR "Global Health care information system" OR "GHIS" )

AND

HIS development process: ("HIS" OR "software process improvement" OR "software process enhancement" OR "CMM" OR "CMMI" OR "SPICE" OR "software process enrichment" OR" software process evaluation" OR "software process assessment" OR "software process appraisal") AND ("positive impact" OR "conclusive" OR "absolute" OR "concrete" OR "perfect" OR "reliable" OR "good effect" OR "best results")

(RQ2) What are the barriers reported in the literature, does it really have negative effect on HIS process Implementation?

**Search String for Barriers**

Barriers: ("barriers" OR "obstacles" OR "hurdles" OR "risks" OR "risk analysis" OR "critical barriers" OR "difficulties" OR "impediments") AND ("GSD" OR "global software development" OR "global software engineering" OR "distributed software development" OR "software outsourcing" OR "offshore software development" OR "information technology outsourcing" OR "software contracting-out" OR "IT outsourcing") AND ("HIS" OR "software process improvement" OR "software process enhancement" OR "CMM" OR "CMMI" OR "SPICE" OR "software process enrichment" OR" software process evaluation" OR "software process assessment" OR "software process appraisal") AND ("negative impact" OR "negative effects" OR "bad impact" OR "bad effects" OR "unwilling impact" OR "unreliable outcomes" OR "undesired results" OR "unsatisfactory results" OR "Global Health care information system" OR "GHIS")

(RQ3) What are the best practices that are reported in the primary studies to address the critical factors and critical barriers?

**Search String for Best Practices:**

The keywords and their alternatives were based on existing literature in the domain of SRE-HRIMM and GSD (Babar, & Niazi, 2008; Ramasubbu, 2014; Khan et al., 2011; Khan, & Keung, 2016, Akbar et al 2020, Shafi et al 2020, Rafi et al 2020, Hamza et al 2020, Tyabba et al 2020, ).

Practices: ("practices" OR "recommendations" OR" implementation practices" OR" improvements" OR "best practices" OR "solutions" OR "exercise" OR "suggestions" OR "execution") AND ("critical factors" OR "important factors" OR "success factors" OR "positive elements" OR "critical success factors" OR "significant variables" OR "significant positive factors") AND ("critical barriers" OR "significant barriers" OR "important obstacles" OR "key barriers" OR "important barriers" OR "critical risks" OR "difficulties" OR "impediments") AND ("GHIS" OR "global software development" OR "global software engineering" OR "distributed software development" OR "software outsourcing" OR "offshore software development" OR "information technology outsourcing" OR "software contracting-out" OR "IT outsourcing" OR "Global Health care information system" OR "GHIS").

### 3.5.1.3 Inclusion and Exclusion criteria

In this section, I have developed the criteria for inclusion/exclusion of the articles in the final selected primary studies.

- **Inclusion criteria:** The primary studies selected must be available in the English language. Every primary research article must be a meeting, journal, or chapter of a book. I concentrated on the papers that addressed the process improvement initiatives in the GHIS context. Further emphasis was on those papers which discussed the success factors and barriers of the HIS development process along with their practices.

- **Exclusion criteria:** I also removed certain documents which have not identified success factors and obstacles to the progress of the software process. Some papers were also omitted that did not include specific information on process management requirements and models. Duplicate articles were also not considered. In addition, certain articles which have been published in other languages besides English have also been omitted.

### 3.5.1.4 Data sources

After identifying keywords and their alternatives, different repositories were chosen to classify the related documents, journals, conference papers, etc. Databases have been selected based on previous research experience, personal knowledge, preferences or suggestions from other researchers (Chen, Babar, & Zhang, 2010, Akbar et al 2020, Nassrullah et al 2019). Data sources shall include:

A. IEEE Explore
B. ACM Digital Library
C. Springer Link
D. Wiley Inter-Science
E. Science Direct
F. Google Scholar

### 3.5.1.5 Quality Criteria for studies selection

Quality assessment (QA) of the selected papers was carried out at the same time as the data extraction process. The guidance provided by (Petticrew, & Helen, 2008; Crombie, & Davies, 1997, Akbar et al 2019, Ahmade et al 2019, Shafiq et al 2019, Tanveer et al 2019, Hamza et al 2019, Badour et al 2019) was followed in the design of the checklist shown in Table 3.2. For. QA1 to QA5 object, the evaluation was carried out as follows:

A. 1 point was assigned to the articles containing the answers to the questions on the checklist.
B. 0.5 points were allocated to the articles containing partial answers to the checklist problem.
C. 0 points were assigned to the articles which did not contain any answers to the questions on the checklist.
The selected studies were also assessed through an informal external review, which confirmed that these studies had sufficient quality to be considered in this SLR study.

Table 3.2 Checklist for Quality assessment of selected studies

| S.No | Checklist Questions |
|---|---|
| QA1 | Do the selected methodologies have addressed the research questions? |
| QA2 | Does the article have discussed software process improvement success factors or barriers? |
| QA3 | Do the study report practices for the SRE-HIMM success factors or barriers in GSD? |
| QA4 | Are the reported results are related to process improvement? |
| QA5 | Do the results have addressed the research questions? |

### 3.5.2 Conducting the review

### 3.5.2.1 Primary studies selection

Various research papers have been found during the primary study selection process and a tollgate method suggested by (Afzal, Torkar, & Feldt, 2009, Akbar et al 2019, Akbar et al 2018, Ahmed et al 2018, Mateen et al 2018, Shafiq et al 2018, Jun et al 2019, Akbar et al 2017) has been used to improve the selection process. As can be seen from Figure 3.2 and Table 3.3, the tollgate method consists of five phases of Ph-1 to Ph-5:

- Phase 1 (Ph-1): consider search term to explore.
- Phase 2 (Ph-2): Title and abstract based inclusion/exclusion.
- Phase 3 (Ph-3): Introduction and conclusion-based inclusion/exclusion.
- Phase 4 (Ph-4): Full text-based inclusion/exclusion.
- Phase 5 (Ph-5): final selection of the articles to be included in the primary studies selection.

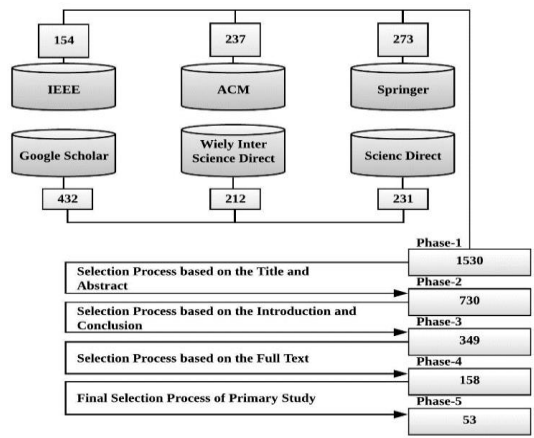

Figure 3.2 Tollgate approach for primary studies selection

Originally, a total of 1530 publications were retrieved from selected repositories on the basis of inclusion and exclusion criteria (section 3.5.1.3) and stage search strings (section 3.5.1.2). The tollgate method (Afzal, Torkar, & Feldt, 2009) contributed to a shortlist of 53 remaining papers to be included in the primary studies. Ultimately, the chosen papers were analyzed on the basis of the quality assurance criteria (section 3.5.1.5).

Table 3.3 Total selected primary studies using tollgate approach

| Electronic databases | Phase-1 | Phase -2 | Phase-3 | Phase-4 | Phase-5 | % of final selected studies (N=53) |
|---|---|---|---|---|---|---|
| ACM Digital Library | 237 | 107 | 48 | 13 | 6 | 11 |
| IEEE Xplore | 154 | 78 | 75 | 35 | 9 | 17 |
| Wiley Inter Science | 212 | 164 | 91 | 39 | 7 | 13 |
| Springer Link | 273 | 128 | 85 | 41 | 8 | 15 |
| Science Direct | 231 | 124 | 78 | 23 | 6 | 12 |
| Google Scholar | 432 | 321 | 209 | 137 | 17 | 32 |
| Total | 1530 | 922 | 586 | 288 | 53 | 100 |

### 3.5.2.2 Data Extraction

To answer the research questions, I have extracted the following data from each article: title, year, study type (journal, conference, thesis, research report), reference, success factors for faced by global health care information system, barriers for faced by healthcare information system, practices to address the critical success factors and critical barriers, research method (QS= Questionnaire Survey, INT= Interview, SLR= Systematic literature review, ILR= Informal literature review, CS= Case Study, MM=Mixed Method, AR=Action Research), continent (AS= Asia, EU= Europe, AUS= Australia, NA= North America, SA= South America, MX=Mixed), company size (small, medium, large).

### 3.5.2.3 Data Synthesis

The author has carried out a data synthesis process. As a result of the extraction phase, a list of success factors and barriers to the remaining 53 articles (53 selected primary studies) was drawn up. Results from all selected articles were reported together and all research questions were evaluated against these results.

### 3.5.3 Reporting the review

### 3.5.3.1 Quality attributes

Appendix B presents the list of scores for each of the listed studies against the five QA issues mentioned in Table 3.2. The final QA score for each study selected was provided by adding scores for all QA questions. All of the 53 studies selected were those that offered sufficient information on the success factors and obstacles of the Global Health Information System (GHIS) and thus 75% of the selected studies scored 50% or more.

### 3.5.4 Empirical data collection

Researchers have a broad selection of empirical research techniques from which they can select the most suitable one to remedy their studies problems. The selection of empirical research strategies ought to be based totally on the sort of facts, to be had resources, control over the chosen approach, and the functionality to perform variables of interest (Bryman, 2012). The complete empirical studies layout of this observe is proven in Figure 3.3, which consists of numerous steps involved in the choice of the targeted populace and respondents, sample layout, instrument improvement and statistics analysis. These steps are briefly explained inside the following sections.

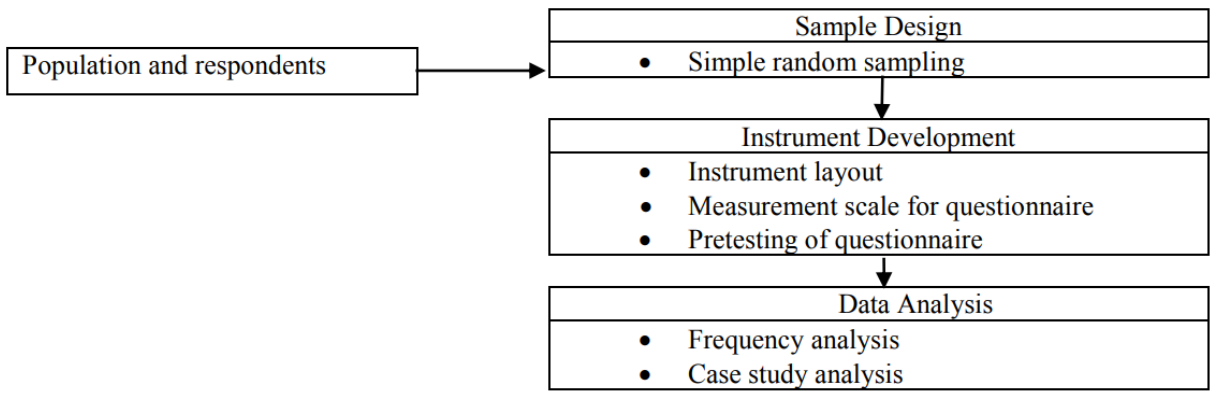

Figure 3.3 Empirical research design process

### 3.5.4.1 Population and Respondents

It is necessary to specifically define the target group in empirical studies. The population is a collection of variables that researchers want to conclude (Borrego, Douglas, & Amelink, 2009). The survey analysis methodology offers an incentive to choose the right sample size as it is difficult to obtain data from the entire population (Borrego, Douglas, & Amelink, 2009). Since the purpose of this empiric analysis is to gain an understanding of the variables that have a positive or negative effect on the global health information system. As GHIS is part of this research analysis, I will need to gather data from a wide variety of Global experts involved in the GSD project.

### 3.5.4.2 Sampling design

Sampling is a method in which appropriate numbers of products are collected from the population (Sekaranm & Bougie, 2010). The sampling approach provides various techniques for gathering.

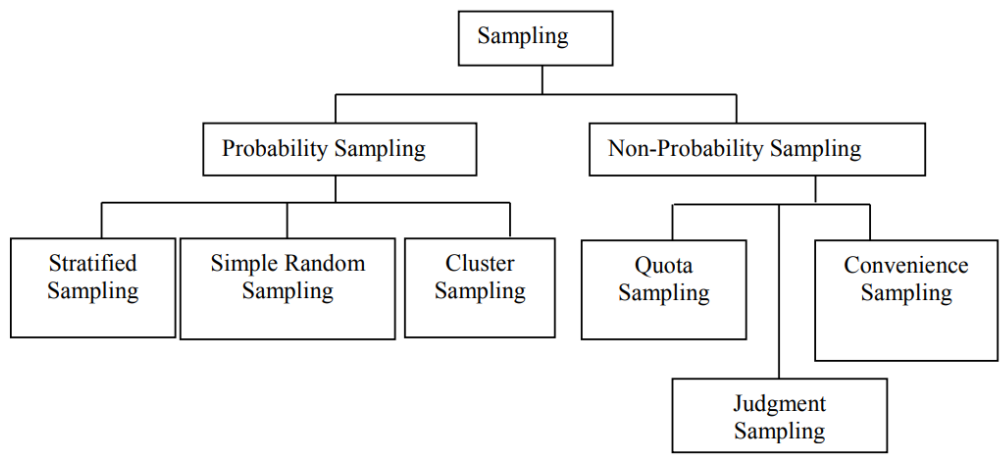

Figure 3.4 Sampling techniques

data from the target population. There are basically two sampling techniques: probability and non-probability of sampling (Malhotra, 1996). Probability sampling may be further categorized as stratified sampling, simple random sampling, and cluster sampling. Likewise, non-probability sampling can be categorized as quota sampling, convenience sampling, and decision sampling (Malhotra,1996).

In this study, an easy random sampling technique is used. Generally, random sampling is used when the quantitative studies method is used to gather the data (Bryman, 2012). In random sampling, each variable of the target population has an equal threat of selection within the sample (Bryman, 2012). If the population is large, random sampling is the first-rate manner to acquire a proper sample of the populace (Fraenkel, Wallen, & Hyum, 2011, Mateen et al 2016, ). It is a right selection technique for the listed members of the population (Lee, & Lings, 2008). As the target population of this take a look at was large, because it consisted of a various variety of GSD groups, consequently, the easy random sampling technique turned into appropriate. The proposed version for this work become intended to use to all GHIS companies approximately generalization; the easy random pattern method is mostly used in such cases. It was not easy to approach the target population as I did now not have the touch information of the experts and also did no longer discover any online repository which has such information. Based on our experience, and discussions with studies colleagues and professionals at Nanjing University of Aeronautics and Astronautics, I determined to approach the target populace the use of extraordinary on-line assets through several agencies hosted by way of LinkedIn (www.Linkedin.Com) and Facebook (www.Facebook.Com). These corporations supplied a possibility for the information device and the GHIS communities internationally to alternate views, ideas, and information related to developing trends. We extensively utilized email to contact the healthcare information machine and GIS specialists. Besides that, I have used an e-mail method to contact the healthcare data machine and GSD professionals. An on-line request was published on the social media businesses, and separate emails were dispatched to certain specialists to invite them to participate. I have contacted 341 members of which 111 finished the questionnaire. I even have manually reviewed all responses to exclude incomplete entries. However, I did now not locate any incomplete responses. We have collected records from specific continents and the bulk of the groups were from Asia and Europe. The respondents ranged from software developers to CEOs with know-how in SRE-HIMM and GHIS. The whole information of the respondents is given in Appendix D

### 3.5.4.3 Survey instrument development

The creation of the data collection instrument consists of different steps and processes to refine the final questionnaire. A closed and open-ended questionnaire for the survey was created. The following steps have been taken to improve the survey instrument.

• **Instrument layout**: The creation of the survey method is an important activity. Each part of the questionnaire should be built on the basis of research questions and research objectives. Other important aspects of the development of the questionnaire are the organization of all questions in such a way as to make it easier for respondents to read and understand (Rea, & Parker, 2014). The questionnaire for this study is divided into four parts, i.e. It's A, B, and C.

Section-A1 includes the questions associated with the respondent's general information. In this phase, the questions have been designed to gather records concerning respondent contact details, position, and experience in SRE-HIMM and GHIS projects.

Section-A2 entails questions designed to gather facts regarding the primary enterprise of the organization, size of the company, a wide variety of employees, type and followed SRE-HIMM models, and standards.

In section-B1 a closed-ended checklist of the fulfillment factors was supplied to the respondents. These achievement elements had been extracted from the literature on the use of the SLR approach. In this segment, the respondents had been requested to specify those success elements that could positively effect SRE-HIMM in GSD ranged from extraordinarily agree to extraordinarily disagree. This section also incorporates open-ended questions to identify additional new achievement elements inside the industry.

In section-B2 a closed-ended tick list of boundaries was supplied to the respondents and they had been asked to specify limitations that could negatively affect SRE-HIMM activities in GHIS ranged from extremely agree to extraordinarily disagree. I have also added an open-ended text field to identify extra new obstacles inside the industry.

Section-C1 entails practices to cope with the recognized important success elements mentioned in phase B1. A total of six success elements had been ranked as vital and a tick list of various practices was furnished for each factor. The respondents were requested to rank every practice for the specific critical fulfillment factor, ranged from extremely agree to extremely disagree.

Section-C2 is the last part of the survey instrument. In this phase, a tick list of practices was supplied to the experts to rank each practice for the recognized essential barriers. A total of 6 barriers was diagnosed as the most crucial obstacles and one of a kind practices mentioned for every barrier.

**The measurement scale for the questionnaire:** I have decided on a seven-point Likert scale for the survey instrument. Researchers have noted diverse reasons for the usage of the seven-factor Likert scale. The downside of the neutral against the seven-point Likert scale is a "myth" and most of the researchers trust that it's miles desirable (Finstad, 2010). Moreover, it's miles comprehensible that respondents might have a neutral opinion concerning a statement or particular subject (Rea, & Parker, 2012). The unavailability of impartial answers in survey questionnaires could compel the respondents to pick either a tremendous or a negative option in response to every question, which can induce bias inside the results. The seven-factor Likert scale gives more correct measures of respondent's true evaluation (Finstad, 2010). The seven-point Likert scale is more appropriate for online survey statistics collection (Finstad, 2010). The survey questionnaire sample is exhibited in Appendix E.

**Pretesting of the questionnaire:** It is important to check the survey instrument to determine the validity of its content by measuring significant variables (Willis, Royston, & Bercini, 2006). Pretesting can be categorized into four different stages , i.e. content validity, readability of the questionnaire, pilot study and final review, as shown in Figure 3.5 (Dillman, 2007).

**Content validity:** The validity of the content of the survey instrument is relevant for evaluating the validity of its contents by measuring significant variables (Rea, & Parker, 2012). In this study, the content analysis was conducted with five experts in the field of empiric software engineering at the Department of Computer Science, Nanjing University of Aeronautics and Astronautics. The following concerns were posed by the reviewers (Dillman, 2007).Are the variables of interest properly presented?

- ➢ Does the survey content consist of any ambiguous items?
- ➢ Does the survey instrument have a proper measurement scale?

Participants checked key variables, terminology and scales of measurement. The reviewers recommended changing the measurement scale to a seven-point Likert instead of a five-point Likert. They noted that the 7-point measurement scale provides more accurate results and is suitable for online questionnaire surveys. During this study, all suggested improvements were made to the survey questionnaire.

**Questionnaire readability:** Various experts were selected to check the wording and clarity of the elements and variables discussed in the questionnaire. These experts included co-supervisor researcher, associate professor of software engineering at the Information and Computer Science Department, King Fahd University of Petroleum and Minerals, Saudi Arabia, researcher supervisor who is an expert in empiric software engineering, and two doctoral students working in the Software Engineering Research Group, City University of Hong Kong. Experts asked the following questions (Dillman, 2007)Are the survey questions properly understood?

- ➢ Can the survey respondents properly interpret the survey questions?
- ➢ Can the survey respondents read and answer the survey questions?
- ➢ Is the survey instrument structured professionally?

After assessment from the experts, all their comments and suggestions were noted and modified the questionnaire accordingly.

**Pilot study:**

A pilot project is a trial project that is performed prior to the finalization of a research design to help identify research problems or to test the viability, reliability and validity of the proposed design of the study (Bernard, 2011). This is a small-scale analysis conducted to test the nature and process of a research sample (Thabane et al., 2010). A pilot study was carried out to check and revise the reliability and accuracy of the instruments from the point of view of the respondent.

- ➢ I have checked the following information at this stage (Rea, & Parker, 2012):Responses were checked whether the answers are uniformly distributed among the respondents.
- ➢ Each response was evaluated to check any question that has been missed by the respondents.
- ➢ The instrument was evaluated in subjects that could harm the response rate.
- ➢ Correlation analysis was performed to check the multi-collinearity.

The results of the pilot studies assist to improve the quality of the survey instrument (Dillman, 2007). I have modified the survey instrument based on the comments and suggestions reported during the pilot study. The summarized comments are given below.

- ➢ Most of the comments were related to the design of the instrument and the respondents suggested designing the questionnaire in the tabular form for more clarity and interest.
- ➢ The second most was respondent 's complaint about the length of some questions.
- ➢ The regression analysis test was employed between the proposed variables and factors.

**Final review:** The survey instrument was finalized based on the final review of my supervisor and co-supervisor. The survey questionnaire was finalized after final review feedback and minor corrections were made in the instrument for the final version.

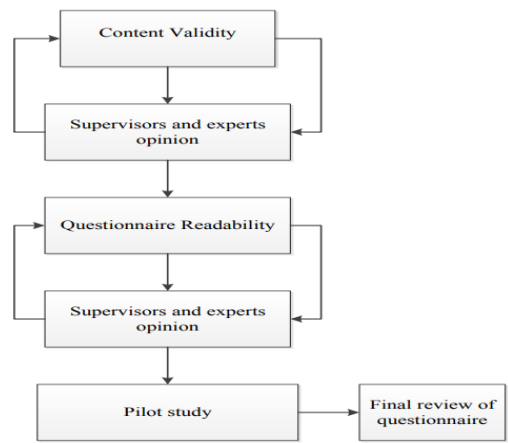

Figure 3.5 Pretesting of Survey Instrument

## 3.6 Data analysis

The researcher used the following research methods to analyze the collected data.

### 3.6.1 Frequency Analysis

Frequency evaluation shows the range of observations for every variable. Frequency evaluation is useful within organizations of variables or throughout them (Khan, 2011, Akbar et al 2020). It may be used to investigate nominal, ordinal, and numerical records (Khan, 2011, Ali, & Khan, 2016, Akbar et al 2020). In this research, each of the SLR and survey questionnaire facts was analyzed using frequency evaluation because it facilitates examine descriptive information (Khan, 2011, Akbar et al 2020). I even have followed the frequency analysis to calculate the occurrences and comparative analysis of each achievement thing and barrier. The comparative analysis of the diagnosed success elements calculated the relative importance of each element. Similarly, the comparative analysis of the diagnosed boundaries highlighted the relative importance of each barrier. Coding is used to change qualitative statistics into quantitative information for analysis purposes (Seaman, 1999; Khan, 2011; Niazi, 2004, Akbar et al 2020). In this research, the data collected the use of an SLR, and survey questionnaires are categorized using a coding technique to carry out the frequency and comparative evaluation of the identified fulfillment factors and barriers. I have calculated the frequency of every component identified in the use of the SLR process. To compute the relative importance of a selected aspect in numerous papers, I even have in comparison the frequency of that specific element against different other elements in the identical number of papers. For example, if an issue has frequency 26 out of 53 then it has 50% importance as compared to other elements. I observed this frequency evaluation method to rank all the identified factors. Identically, I have calculated the frequencies of the recognized factors in all the accumulated survey questionnaires. I even have done comparative evaluation for each issue via evaluating the frequency of a specific issue in all the survey questionnaires towards the frequencies of different factors in the equal wide variety of survey questionnaires (Niazi, 2004; Khan, 2011; Ali, & Khan, 2016, Akbar et al 2020). The same procedures of frequency analysis and coding are used by extraordinary other researchers for the equal nature of the data (Khan, 2011; Niazi, 2004; Ali, & Khan, 2016, Akbar et al 2020).

### 3.6.2 Case study for SRE-HIMM evaluation

The case study approach was introduced to test the proposed model (SRE-HIMM) as it is considered to be the most common evaluation methodology (Yin, 2009). A total of three case studies have been performed in three separate GSD organizations to test SRE-HIMM (see Chapter 6). A feedback session was also conducted with the case study participants to receive input on the features of the SRE-HIMM. The following parameters have been established to gain input from the participants:

- ➢ Ease of use
- ➢ User satisfaction
- ➢ Structure of SRE-HIMM

The above principles can be used to evaluate and measure the consistency and efficacy of goods and can help to identify areas with defects (Khan, 2011). Such parameters were based on studies performed by various other scholars in different other fields (Khan et al., 2011; Khan, 2011; Niazi, 2004; Niazi et al., 2004; Niazi, Wilson, & Zowghi, 2005; Khan, 2015; Ali, & Khan, 2016; Akbar et al 2020).

## 3.7 Threats to Validity with research methodologies

### 3.7.1 Threats to validity (SLR)

There are several drawbacks to systematic literature review studies. One potential drawback is that most of the SLR data were obtained by the author of this research study and, due to a large number of research papers on SRE-HIMM and GSD, this research study may have overlooked some related posts. However, this is not a systemic omission like other SLR researchers (Khan & Khan, 2013). For selected primary studies, the possible drawback is that their identified success factors and obstacles may not have been identified for the primary reason. It's a challenge for me to ease this threat. The authors have not discussed the basic reasons why the success factors and barriers were considered.

### 3.7.2 Threats to Validity (Empirical Study)

In this study's work, I actually have explored the views and opinions of experts regarding SRE-HIMM implementation in GHIS organizations. I was now not able to directly verify the views and reviews of the experts. It approaches that the remarks from experts regarding the factors might not necessarily be authentic.

Construct validity represents whether or no longer the measurement scales denote the attributes being measured. The elements taken into consideration in this have a look at had been received from an extensive body of literature suggested in (Babar, & Niazi, 2008; Niazi, Babar, & Verner, 2010; Niazi, 2015, Akbar et al 2019) and dialogue with the SRE-HIMM practitioners. The remarks from practitioners demonstrates that each one the chosen elements were associated with their work. Internal validity refers to the general evaluation of the results. The consequences of the pilot look at furnished an acceptable level of internal validity as the elements considered in this research work had been acquired from an in-depth literature evaluation and piloting of questions. External validity refers to the generalization of the results for all different domains (Regnell, Runeson, & Thelin, 2000, Akbar et al 2019). In this have a look at, I was no longer able to generalize our have a look at outcomes as maximum of the respondents were from Asian nations. We can't claim that each one the respondents from the selected countries would trust them, however we are certain that they furnished a consultant sample.

In this survey questionnaire, the experts had been furnished close-ended inquiries to rank the identified fulfillment factors and barriers together with their practices. The close-ended questions restrict the survey respondents to best those success factors, barriers, and practices which might be provided inside the list. I attempted to eliminate this problem by providing an open-ended text field to the respondents to say additional elements besides the mentioned factors. However, like other researchers (Baddoo, & Hall, 2002; Beecham, Hall, & Rainer, 2003; Niazi, Wilson,

& Zowghi, 2006, Khan, 2011, Akbar et al 2018), I believe on the accrued responses due to the fact the experts operating in diverse roles, obligations and directly worried in software program process development and GHIS activities.

**3.8 Summary**

This chapter presented detailed research methods, which were briefly described. The chapter offered a detailed description of the SLR process. The SLR addressed the stages of the preparation of the review, the execution of the review and the reporting of the review. A brief study design was addressed for the empirical research process, with their conceptual references suggested where applicable. In addition, the reasons for selecting the target population, sampling, instrument development and pre-testing, the data collection approach, the data collection questionnaire and the brief data analysis method have been explained in detail. In the end, the weaknesses of the chosen study methodologies have been identified.

# Chapter 4: SLR AND EMPIRICAL STUDY RESULTS

## 4.1 Overview

This chapter comprised of results attained by SLR and empirical study conducted using the survey questionnaire approach. The main objective of this study is to address success factors, barriers, their practices, and discussed the most Critical Success Factors and Critical Barriers (CBs). The CSFs and CBs present some of the key areas related to SRE-HIMM implementation and the understanding of CSFs and CBs could assist GSD Healthcare organizations to successfully execute SRE-HIMM activities.

The ultimate aim of this research study is to develop a software requirement engineering Healthcare implementation Maturity model (SRE-HIMM) (Chapter 5).

This chapter is organized as follows:

- In Section 4.2, the success factors identified using an SLR and survey study is discussed.
- In Section 4.3, the barriers identified using an SLR and survey study is discussed.
- In Section 4.4 the practices of CSFs and CBs are reported.
- Sections 4.5 consist of the results and analysis of the identified success factors and barriers discussed in various aspects.
- The summary of the chapter is given in section 4.6

## 4.2 Identified success factors using SLR and survey questionnaire

This section consists of the success factors found by the SLR and the survey techniques. The findings discussed in this section answered the following research questions in this report.

RQ1. What are the factors reported in the primary literature, do GHIS organizations need to address reported factors for the successful implementation of HIS process?

### 4.2.1 Success factors identified using SLR

To address the RQ1, a list of success factors identified using the SLR approach is shown in Table 4.1 and Table 4.1. The identified factors have been extracted from the list of 53 selected primary studies. The most frequent success factors were selected with the original wording and the others were paraphrased in order to reduce the similarity index of the factors. For example, SF1: user and top management commitment and SF2: resource allocation was stated as discussed in the literature.

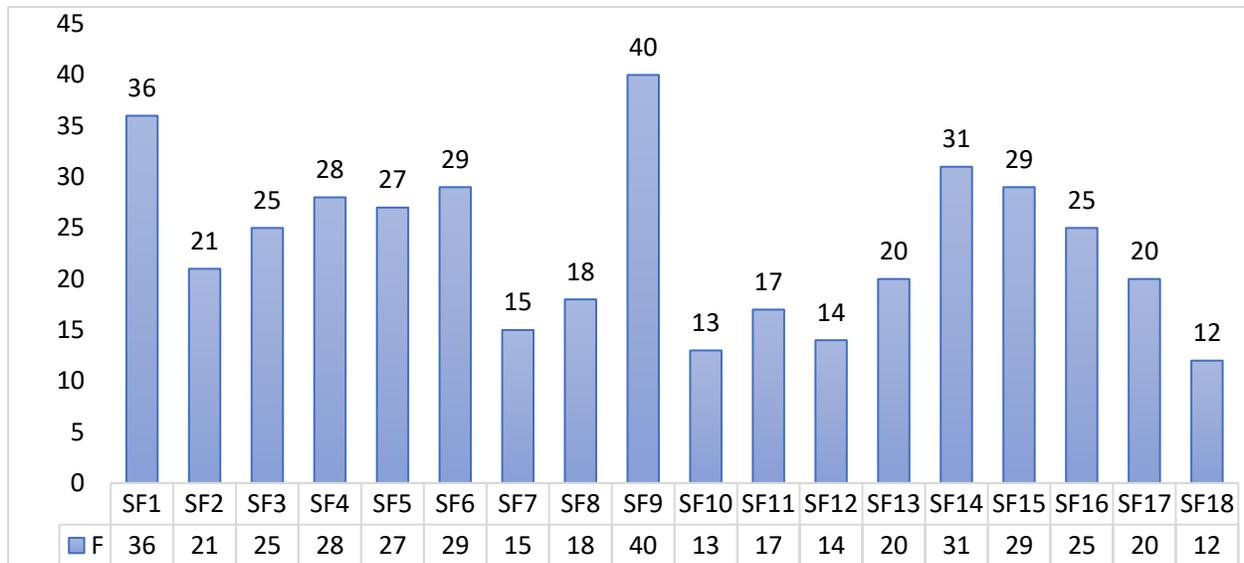

Figure 4.1: Frequency distribution of Success Factors identified using SLR

In Figure 4.1, (N =53) is the total number of articles selected for the primary study and these articles are cited in Appendix B. Each of the selected primary study labelled as [LT], which aims to present the primary studies of SLR as the "Literature Review" articles, as shown in Appendix B.

Table 4.1 Success factors identified through literature

| S.No | Identified success factors | Rank | S.No | Identified success factors | Rank |
|------|----------------------------|------|------|----------------------------|------|
| SF1  | User's and Top Management commitments | 2 | SF9 | 3C's (communication, coordination, and control) | 1 |
| SF2  | HIS skilled labor | 8 | SF10 | Project Pilot Implementation | 14 |
| SF3  | Role of Collective Learning | 7 | SF11 | Resource allocation | 11 |
| SF4  | Protect Patient Privacy and Data Security | 5 | SF12 | Team collaboration | 13 |
| SF5  | Staff training and involvement | 6 | SF13 | Organizational culture and contextual | 9 |
| SF6  | Process improvement leadership | 4 | SF14 | The strong relationship between team members | 3 |
| SF7  | Hospital-wide process Improvement | 12 | SF15 | Process Improvement Protocols and Procedures | 4 |
| SF8  | Interdisciplinary team communication | 10 | SF16 | Involving End Users | 7 |
| SF17 | Assess capacity and capabilities within an organization | 9 | SF18 | Data Quality Improvement | 15 |

We have identified eighteen success factors that have a positive influence while implementing HIS development. We further have noticed that seven factors that are presented in literate have frequency ≥ 50.

Therefore, among the eighteen success factors identified, 'SF9: 3C's (communication, coordination, and control)' was reported (75%) and is the most common factor in selected primary studies as a success factor for HIS implementation. Kuziemsky et al. [LT18] argued that communication between members of the health care team is essential to the implementation of the project. Communication is the process of information transfer between the dispersed team members and the mode of communication they follow to facilitate this interaction. Active contact networks are believed to help boost the process (Weller et al., [LT21]). Good communication eliminates the noise, convinces team members of the need for process change, and decreases their resistance to change (Weller et al., [LT21]). Coordination refers to the involvement of various individuals working together on a common organizational mission (Kuziemsky et al. [LT18]). Coordination makes team members interdependent. When the team members are working on the same project but in different directions than the lack of teamwork (Taweel et al., [LT52]).

SF1: User's and Top Management commitments' (68%) is the second most common success factor reported in the literature for HIS Process implementation as in Table 4.1. According to (Cardoso et al., LT1), User's and Top Management commitments are the extents to which the higher and lower level management in organization support, finance, realize and participate in the HIS Implementation program. Cardoso et al. [LT1] identified User's and Top Management commitments as a positive influential factor for HIS implementation.

SLR results exhibit that 'SF15: Process improvement procedure and protocol' (54%) is reported as the critical factor that has a strong influence on the development of HIS development. It is reported that the protocols and procedures play a vital role in the development of HIS. Organizations should develop quantifiable protocols and procedures so that they could manage everything smoothly specified that the leadership of the HIS team is vital for the successful execution of the HIS program. The management should have the ability to direct and encourage team members to achieve specific goals (Taweel et al., [LT52]. If the process manager has good awareness and adequate skills, there is an increased possibility of HIS success and acceptance in the best practice of domain (Taweel et al., [LT52]).

Similarly, SF5: Staff Training and Involvement (50%) is reported in the Literature, which ensures that the involvement of health team members is likely to play an important role in the process of raising the skills of the workforce, which has been given a central role in the economic strategies as well as the development of HIS systems in many nation-states. Staff training plays a vital role in the development of the HIS system. Enhancing their expertise and participation in the creation phase of the HIS method would benefit. Weaver et al. [LT2] figured out the twelve best practices that could improve staff training and encourage them to participate in team activities. Ajami et al. [LT3] addressed the training of team members and its positive impact on the successful implementation of the HIS development Process.

The results of the SLR show that 'SF14: the strong relationship between team members' (58%) is a key factor in the successful implementation of the HIS process. This is the extent to which members of the HIS team could effectively communicate and collaborate in the implementation of the HIS program (Ortner et al., [LT9]). According to (Singer et al, [LT10]), a good partnership between the dispersed teams may help team members identify new HIS policies, procedures, and strengthen existing ones. He also concluded that not only the management staff but also the lower-level staff should have a complete understanding of the strong relationship between the members of the team. HIS production cycle without support and confidence, Distributed process improvement teams may not be able to complete HIS tasks within budget and time. Top management would empower dispersed team members to be positive and eager to engage in related HIS activities.

In addition, SF4: Protect Patient Privacy and Data Security (52%) is frequently reported in the literature to ensure the protection of the private patient record to be stolen by unauthorized persons. Doctors must be responsible for helping to protect electronic medical information. They must document all use of user information, share their privacy and security policies with you, and report any loss of information. Contact your doctor's office immediately if you suspect someone has misused your electronic health information.

Furthermore, HIS skilled labor, Role of collective learning, Hospital-wide process improvement, project pilot implementation, Resource allocation, interdisciplinary team communication, process improvement protocols and procedures, Strong relationship between team members, Team collaboration, were also cited as the success factors that have a positive impact on SRE- HIMM Improvement.

### 4.2.2 Success factors identified using a survey questionnaire

In order to address RQ3, Table 4.2 shows the list of success factors identified during the empirical study conducted in GHIS organizations. A total of 77 responses was collected from a total of 14 countries. Table 4.2 is divided into three main categories of 'Positive' (Strongly Agree (SA), Moderately Agree (MA), 'Negative' (Strongly Disagree (SD), Moderately Disagree (MD)) and 'Neutral' (NS). The positive category from the table presents the percentage of the respondents who agree with the success factors identified using the SLR technique. The negative category shows the percentage of those respondents who consider that the identified success factors might not be effective for HIS implementation in the context of GHIS. In the neutral category, I have mentioned the percentage of those respondents who were not sure about the ranking of the identified factors.

The results are shown in Table 4.2 exhibits that all the factors have a frequency greater than 50% and success factor 'SF1: users' and top management commitment (88%)' is the most common factor in the positive list. Success factors 'SF15: process improvement protocols and procedures (77%)' 'SF7: Hospital-wide process improvement (79%)' are also considered common success factors highlighted by the experts SF6: Process improvement leadership (86%)is the 2nd most important factor in the positive list of the responses. Other common factors are 'SF5: Staff training and involvement (83%).

Table 4.2 Success factors identified through a survey questionnaire

| S.No | Success Factors | Experts Opinion (N=77) | | | | | | | |
|---|---|---|---|---|---|---|---|---|---|
| | | +Ve | | | -Ve | | | Neutral | |
| | | EA | MA | % | ED | MD | % | NU | % |
| 1 | User's and Top Management commitments | 41 | 27 | 88 | 4 | 2 | 7 | 3 | 4 |
| 2 | HIS skilled labor | 30 | 26 | 80 | 7 | 3 | 13 | 5 | 6 |
| 3 | Role of Collective Learning | 37 | 28 | 84 | 6 | 4 | 13 | 2 | 2 |
| 4 | Protect Patient Privacy and Data Security | 32 | 24 | 72 | 4 | 3 | 9 | 14 | 18 |
| 5 | Staff training and involvement | 39 | 25 | 83 | 7 | 2 | 11 | 4 | 5 |
| 6 | Process improvement leadership | 39 | 27 | 86 | 4 | 2 | 7 | 5 | 6 |
| 7 | Hospital wide process Improvement | 37 | 24 | 79 | 7 | 4 | 14 | 5 | 6 |
| 8 | Interdisciplinary team communication | 35 | 23 | 75 | 8 | 5 | 16 | 6 | 7 |
| 9 | 3C's (communication, coordination, and control) | 41 | 27 | 89 | 4 | 2 | 7 | 3 | 3 |
| 10 | Project Pilot Implementation | 28 | 27 | 71 | 9 | 7 | 20 | 9 | 11 |
| 11 | Resource allocation | 30 | 22 | 67 | 9 | 7 | 5 | 5 | 4 |
| 12 | Team collaboration | 39 | 25 | 84 | 7 | 2 | 11 | 4 | 5 |
| 13 | Organizational culture | 34 | 26 | 77 | 7 | 4 | 14 | 6 | 7 |
| 14 | Strong relationship between team members | 38 | 24 | 83 | 7 | 2 | 11 | 4 | 5 |
| 15 | Process Improvement Protocols and Procedures | 34 | 26 | 77 | 7 | 4 | 14 | 6 | 7 |
| 16 | Involving End Users | 35 | 24 | 76 | 9 | 3 | 15 | 6 | 7 |
| 17 | Assess capacity and capabilities within organization | 29 | 19 | 62 | 11 | 5 | 20 | 13 | 17 |
| 18 | Data quality Improvement | 29 | 23 | 67 | 9 | 7 | 20 | 9 | 11 |

In the negative list of the responses, 'SF10: Project pilot implementation (20%)' is the most common factor as shown in Table 4.2. It shows that 20% of the survey respondents declined 'SF17: Assess capacity and capabilities within the organization' as the success factor for HIS implementation.

### 4.2.3 Comparison of success factors identified using SLR and survey questionnaire

In this section, a comparison of success factors from SLR and survey questionnaire has been investigated.

The main objective of this comparison is to get insight into the similarities and differences between the two data sets. Table 4.3 shows the summary of all factors identified in the SLR and the questionnaire based on their average rank. I also performed an independent t-test to evaluate the mean difference in ranks of both SLR and empirical studies as shown in Table 4.4 and Table 4.5.

Table 4.3 Comparison of success factors across S LR and empirical study.

| S.No | Success Factors | Occurrence in SLR (n=53) | | Positive occurrence in survey (n=77) | | Average Rank |
|---|---|---|---|---|---|---|
| | | % | Rank | % | Rank | |
| 1 | Users' and Top Management commitment | 68 | 2 | 88 | 2 | 2 |
| 2 | HIS skilled labor | 39 | 8 | 80 | 6 | 7 |
| 3 | Role of Collective Learning | 47 | 7 | 84 | 4 | 6 |
| 4 | Sharing and Communication | 52 | 5 | 72 | 11 | 8 |
| 5 | Staff training and involvement | 50 | 6 | 83 | 5 | 6 |
| 6 | Process improvement leadership | 54 | 4 | 86 | 3 | 4 |
| 7 | Hospital wide process Improvement | 28 | 12 | 79 | 7 | 10 |
| 8 | Interdisciplinary team communication | 34 | 10 | 75 | 10 | 10 |
| 9 | 3C's (communication, coordination, and control) | 75 | 1 | 89 | 1 | 1 |
| 10 | Project Pilot Implementation | 24 | 14 | 71 | 12 | 13 |
| 11 | Resource allocation | 32 | 11 | 67 | 13 | 12 |
| 12 | Team collaboration | 26 | 13 | 84 | 4 | 9 |
| 13 | Organizational culture | 37 | 9 | 77 | 8 | 9 |
| 14 | Strong relationship between team members | 58 | 3 | 83 | 5 | 4 |
| 15 | Process Improvement Protocols and Procedures | 54 | 4 | 77 | 8 | 6 |
| 16 | Involving End Users | 47 | 7 | 76 | 9 | 8 |
| 17 | Assess capacity and capabilities within the organization | 37 | 9 | 62 | 14 | 12 |
| 18 | Data quality Improvement | 22 | 15 | 67 | 13 | 14 |

Using Levene's Test, I have calculated the significant difference for the ranks of both SLR and empirical study. In table 4.5 it is shown that the results of the t-test are (t=1.012, p=0.001)

Table 4.4 Group Statistics for success factors

| Group Statistics | | | | | |
|---|---|---|---|---|---|
| | Group | N | Mean | Std. Deviation | Std. Error Mean |
| Rank | SLR | 18 | 4.4444 | 3.48479 | .82137 |
| | Empirical | 18 | 5.3889 | 1.88302 | .44383 |

Table 4.5 Independent Sample t-test for success factors

| | | Levene's Test for Equality of Variances | | t-test for Equality of Means | | | | | | |
|---|---|---|---|---|---|---|---|---|---|---|
| | | F | Sig. | t | df | Sig. (2-tailed) | Mean Difference | Std. Error Difference | 95% Confidence Interval of the Difference | |
| | | | | | | | | | Lower | Upper |
| Rank | Equal variances assumed | 12.291 | .001 | -1.012 | 34 | .319 | -.94444 | .93362 | -2.84178 | .95289 |
| | Equal variances not assumed | | | -1.012 | 26.147 | .321 | -.94444 | .93362 | -2.86300 | .97411 |

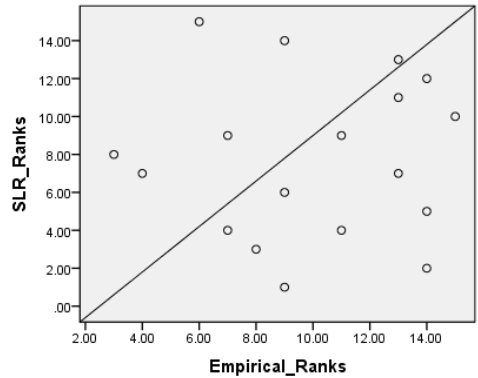

Figure 4.2 Scatterplot of the success factors ranks obtained from SLR and Empirical study

### 4.2.4 Critical success factors identified using both SLR and survey questionnaire

The concept of critical factors was introduced by (Rockart, 1979) to identify the information needs of the chief executive. The concept of critical factors is based on the concept of factors discussed in the literature of management (Daniel, 1961). (Niazi, 2004, Akbar et al 2018) highlighted the concept of critical factors as the key areas where the organizational management must focus to achieve specific business goals. The limited attention given to those key areas can undermine the performance of the business (Khan, & Jacky, 2016, Akbar et al 2018). Critical factors may differ from person to person as it depends on the position of an individual's hold in an organization. It also depends on the geographical regions of the managers and may change with the passage of time (Khandelwal, & Ferguson, 1999; Rockart, 1979, Akbar et al 2018).

I have followed the following criteria in order to conclude the criticality of a specific factor:

- If a factor has a frequency of ≥50% in both literature and empirical study, then the factor is considered as a critical factor.

By using the above criteria, the identified CSFs are User and Top management commitment, staff training involvement, Strong relationship between team members,3C (communication, coordination, and control), Patient Privacy and data security, Process Improvement Protocols and Procedures. The identified CSFs have frequency of ≥50% in both data sets (SLR, empirical study).

## 4.3 Identified barriers using SLR and survey questionnaire

This segment offers a brief overview of the barriers found by the SLR and the empirical research performed in GHIS organizations. The results presented in this section will respond to the following research questions.

RQ2. What are the barriers reported in the literature, does it really have negative effect on HIS process Implementation?

### 4.3.1 Barriers identified using SLR

This segment consists of the results and interpretation of the RQ2 response. Figure 4.4 and Table 4.7 provide a list of obstacles derived from current literature using the SLR method.

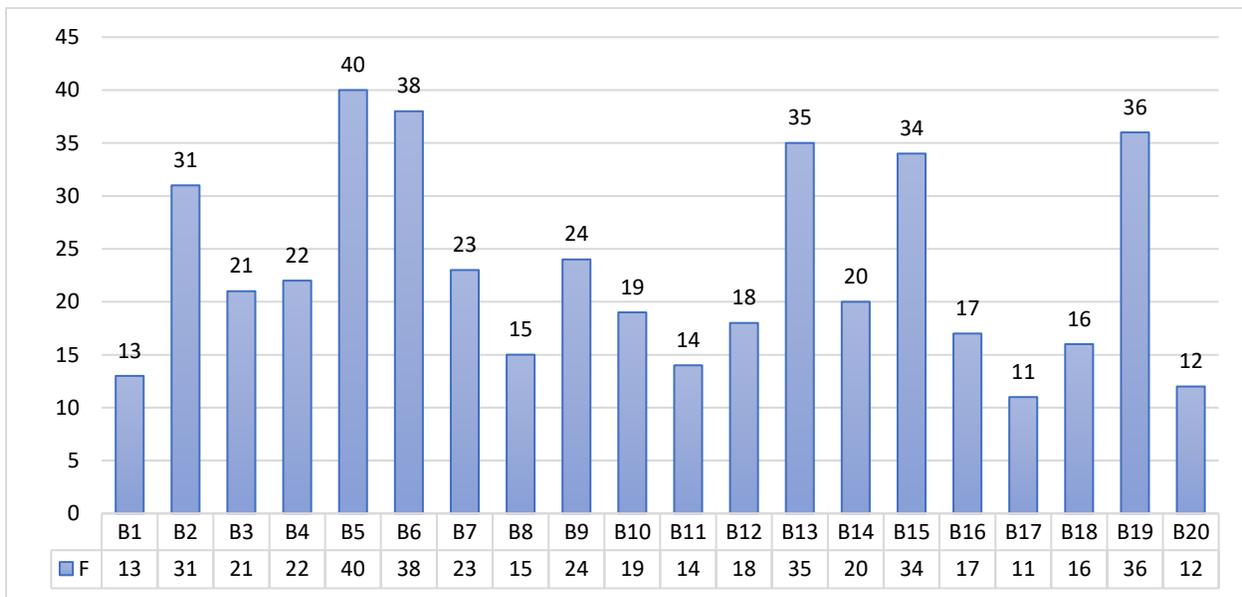

Figure 4.3 Frequency distribution of barriers identified using SLR

The total number of articles selected during the SLR study is shown in Figure 4.4, (N = 53) and these articles are cited in Appendix B. A total of twenty-two barriers were removed from the selected SLR studies as shown in Tables 4.7 and Figure 4.4.

Table 4.7 Identified barriers using SLR

| S.No | Identified barriers | Rank | S.No | Identified barriers | Rank |
|---|---|---|---|---|---|
| B1 | Consider emotions and personal values | 18 | B10 | The negative attitude of society towards using HIS | 12 |
| B2 | Scope change and creep | 6 | B11 | Missing traceability | 17 |
| B3 | Ad-hoc change management and lack of traceability | 10 | B12 | The difference in working situations | 13 |
| B4 | Fear of Intrusion | 9 | B13 | Uncertainties and rapidly changing technology | 4 |
| B5 | Distributed healthcare team management issues | 1 | B14 | Communication flows in project teams and clients | 11 |
| B6 | Lake of Technological Resources | 2 | B15 | Lake of patient engagement | 5 |
| B7 | Un-structured Information | 8 | B16 | Volatile client business domain | 14 |
| B8 | Unclear and fuzzy stakeholders' expectations | 16 | B17 | Regulatory issues between distributed teams | 20 |
| B9 | Lack of terminologies understandings | 7 | B18 | Role of patient addressing security and privacy concern | 15 |
| B19 | National Policies towards HIT | 3 | B20 | Environmental constraints at GHIS sites | 19 |

Among the 20 barriers found, "B5: distributed Health Team Management Challenges (75%)" recorded in the primary studies as there are many challenges that the globally distributed teams face on a daily basis, such as Lack of In-Person Interaction, Time Zone Mismatch, People Come From Different Cultures, The Black Box, Non-Native Speakers. (Bolden et al. [LT22]) discussed the role of leadership in the distributed healthcare system in addressing the key issues facing the distributed environment.

Similarly, "B6: Lack of Technological Resources (71%)" is the second most common barrier that is reported in the primary literature. Modern technological resources play a vital role in the development of complex healthcare Information system. It is envisioned that modern technological resources have a positive impact on the development of any system as it reduces time and employs efforts to provide the most accurate results. Lake of these technological resources may lead to the failure of the entire system.

Similarly, B15: Lake of Patient engagement (64%) stated in the literature that patients do not promote patient engagement. Patients themselves cannot easily access their care records. Online contact with healthcare practitioners and online scheduling of treatments and tests are also restricted (Minister of Health and Social Affairs, 2006). Studies have shown that giving patients more access to their health information will enable them to engage in their own treatment, self-manage their wellbeing, increase awareness of their medical problems, and enhance patient-provider contact (Ricciardi et al., 2013; Delbanco et al., 2012).

Similarly, 'B19: National HIT Policies' (67 %) of SLR articles have been identified as a barrier that could negatively impact HIS activities. Thus, efficient, effective, and secure national policies can address local health needs in a changing environment. Such policies can be formulated by decision-makers and professionals to assess and enforce research evidence. Enforcement of legislation is difficult in developed countries, and societal support for the improvement of any structure is more difficult (Mugdha; Chinnock; Garner et al).

"B2: Scope change and Creep" stated in the literature (58%) that Stakeholder creeps discussion led to an unforeseen increase in time (3-month late go-live) and effort (68% over planned HIT working hours). Stakeholder expansion can have a detrimental effect on the execution of HIT ventures. Projects can be vulnerable to slipping when clinicians/researchers have assumptions about HIT organization and processes. Implementation methods, such as the proposed stakeholder checklist, may help to avoid and minimize the effect of stakeholder creep, Paré, Guy, et al. [LT53].

'B15: The role of patients discussing security and privacy issues (64%) addressed in the literature when devolving the health information system as issues about privacy and protection of electronic health information fall into two general categories : ( 1) concerns about unauthorized releases of information from individual entities and (2) concerns about institutional flows of information. Inappropriate releases from organizations can result either from approved users who deliberately or inadvertently access or disseminate information in violation of organizational policy or from outsiders who break into the organization's computer system. (Meingast et al. [LT42].

B10: The negative attitude of society towards using HIS, B13: Environmental constraints at GHIS sites B12: Difference in working situations, Fear of Intrusion, Ad-hoc change management and lack of traceability, Uncertainties and rapidly changing technology, Un-structured Information, Unclear and fuzzy stakeholders' expectations, The negative attitude of society towards using HIS, Missing traceability, Difference in working situations, Environmental constraints at GHIS sites, Communication flows in project teams and clients,  Volatile client business domain are also the reported in the literature.

### 4.3.2 Barriers identified using a survey questionnaire

A list of identified barriers is given in Table 4.8. The findings indicate that many respondents agreed with the defined obstacles that could have a negative effect on process change in the GHIS climate. This is evident from the results in the 'Positive (+ Ve)' category, where most of the values are above 60% except for a few barriers.

In the identified barriers 'BA15: Lake of patient engagement (88%)' was considered the major barrier for HIS implementation in GHIS. This is because operational management feels time pressure as a preeminent hurdle to process improvement execution. Under time and cost team members made quick decisions in order to stay on schedule. However, those decisions may not be for the benefit of the process improvement program.

The results shown in Table 4.8 highlighted that 'BA15: BA5: Distributed healthcare team management issues (85%)' is the second most common factor for HIS implementation in GHIS.

In the absence of sufficient resources, the organization will not be able to achieve the goals and objectives of the HIS program.

Table 4.8 Barriers identified using a survey questionnaire

| S.No | Success Factors | Expert Opinion =77 | | | | | | | |
|---|---|---|---|---|---|---|---|---|---|
| | | -Ve | | | +Ve | | | Neutral | |
| | | SA | A | % | SD | D | % | NU | % |
| BA1 | Consider emotions and personal values | 31 | 21 | 67 | 11 | 7 | 23 | 7 | 9 |
| BA2 | Scope change and creep | 37 | 24 | 79 | 7 | 4 | 14 | 5 | 6 |
| BA3 | Ad-hoc change management and lack of traceability | 36 | 21 | 74 | 7 | 4 | 14 | 9 | 11 |
| BA4 | Fear of Intrusion | 29 | 19 | 62 | 11 | 5 | 20 | 13 | 17 |
| BA5 | Distributed healthcare team management issues | 39 | 27 | 85 | 4 | 2 | 7 | 5 | 6 |
| BA6 | Lake of Technological Resources | 35 | 24 | 76 | 9 | 3 | 15 | 6 | 7 |
| BA7 | Un-structured Information | 29 | 21 | 64 | 9 | 6 | 19 | 12 | 15 |
| BA8 | Unclear and fuzzy stakeholders' expectations | 29 | 23 | 67 | 9 | 7 | 20 | 9 | 11 |
| BA9 | Lack of terminologies understandings | 33 | 23 | 72 | 12 | 5 | 22 | 4 | 5 |
| BA10 | The negative attitude of society towards using HIS | 27 | 19 | 60 | 11 | 7 | 23 | 13 | 16 |
| BA11 | Missing traceability | 31 | 21 | 67 | 11 | 7 | 23 | 7 | 9 |
| BA12 | Difference in working situations | 31 | 24 | 71 | 9 | 5 | 18 | 8 | 10 |
| BA13 | Uncertainties and rapidly changing technology | 32 | 23 | 75 | 8 | 5 | 16 | 6 | 7 |
| BA14 | Communication flows in project teams and clients | 36 | 26 | 80 | 7 | 3 | 13 | 5 | 6 |
| BA15 | Lake of patient engagement | 41 | 27 | 88 | 4 | 2 | 7 | 3 | 3 |
| BA16 | Volatile client business domain | 33 | 21 | 70 | 11 | 6 | 22 | 6 | 7 |
| BA17 | Regulatory issues between distributed teams | 30 | 22 | 67 | 9 | 6 | 19 | 10 | 12 |
| BA18 | Role of patient addressing security and privacy concern | 33 | 23 | 72 | 12 | 5 | 22 | 4 | 5 |
| BA19 | National Policies towards HIT | 37 | 24 | 79 | 7 | 4 | 14 | 5 | 6 |
| BA20 | Environmental constrains at GHIS sites | 31 | 21 | 67 | 11 | 7 | 23 | 13 | 16 |

**4.3.3 Comparison of barriers identified using SLR and survey questionnaire**

I compared the ranks of the barriers found in both types of studies (SLR and empiric study) as shown in Table 4.9. I also assessed mean differences in SLR and survey data using an independent t-test as shown in Table 4.10 and Table 4.11. The estimated results of the t-test are (t=-1.079, p=0.02>0.05), which indicates that there is no substantial difference between the SLR and the questionnaire findings.

Table 4.9 Comparison of barriers identified using SLR and a survey questionnaire

| S.No | Barriers | Occurrence in SLR (n=53) | | Positive occurrence in survey (n=111) | | Average Rank |
|---|---|---|---|---|---|---|
| | | % | Rank | % | Rank | |
| 1 | Consider emotions and personal values | 24 | 18 | 67 | 10 | 14 |
| 2 | Scope change and creep | 58 | 6 | 79 | 4 | 5 |
| 3 | Ad-hoc change management and lack of traceability | 39 | 10 | 74 | 6 | 8 |
| 4 | Fear of Intrusion | 41 | 9 | 62 | 12 | 11 |
| 5 | Distributed healthcare team management issues | 75 | 1 | 85 | 2 | 2 |
| 6 | Lake of Technological Resources | 71 | 2 | 76 | 5 | 4 |
| 7 | Un-structured Information | 43 | 8 | 64 | 11 | 10 |
| 8 | Unclear and fuzzy stakeholders' expectations | 28 | 16 | 67 | 10 | 13 |
| 9 | Lack of terminologies understandings | 45 | 7 | 72 | 7 | 7 |
| 10 | The negative attitude of society towards using HIS | 35 | 12 | 60 | 13 | 13 |
| 11 | Missing traceability | 26 | 17 | 67 | 10 | 14 |
| 12 | Difference in working situations | 33 | 13 | 71 | 8 | 11 |
| 13 | Uncertainties and rapidly changing technology | 66 | 4 | 75 | 6 | 5 |
| 14 | Communication flows in project teams and clients | 37 | 11 | 80 | 3 | 7 |
| 15 | Lake of patient engagement | 64 | 5 | 88 | 1 | 3 |
| 16 | Volatile client business domain | 32 | 14 | 70 | 9 | 12 |
| 17 | Regulatory issues between distributed teams | 20 | 20 | 67 | 10 | 15 |
| 18 | Role of patient addressing security and privacy concern | 30 | 15 | 72 | 8 | 12 |
| 19 | National Policies towards HIT | 67 | 3 | 79 | 4 | 4 |
| 20 | Environmental constrains at GHIS sites | 22 | 19 | 67 | 10 | 15 |

Table 4.10 Group Statistics for barriers

| Group Statistics | | | | | |
|---|---|---|---|---|---|
| | Group | N | Mean | Std. Deviation | Std. Error Mean |
| Rank | SLR | 20 | 4.3500 | 3.40704 | .76184 |
| | empirical | 20 | 5.3000 | 1.97617 | .44189 |

Table 4.11 Independent Sample t-test for barriers

| Independent Samples Test | | | | | | | | | | |
|---|---|---|---|---|---|---|---|---|---|---|
| | | Levene's Test for Equality of Variances | | t-test for Equality of Means | | | | | | |
| | | F | Sig. | t | df | Sig. (2-tailed) | Mean Difference | Std. Error Difference | 95% Confidence Interval of the Difference | |
| | | | | | | | | | Lower | Upper |
| Rank | Equal variances assumed | 11.003 | .002 | -1.079 | 38 | .288 | -.95000 | .88071 | -2.73291 | .83291 |

| Equal variances not assumed | | | -1.079 | 30.485 | .289 | -.95000 | .88071 | -2.74746 | .84746 |

Table 4.12 Barriers identified using a survey questionnaire

| Correlations | | | SLR_Ranks | Empirical_Ranks |
|---|---|---|---|---|
| SLR_Ranks | Pearson Correlation | | 1 | .58* |
| | Sig. (2-tailed) | | | .046 |
| | N | | 20 | 20 |
| Empirical_Ranks | Pearson Correlation | | .58* | 1 |
| | Sig. (2-tailed) | | .046 | |
| | N | | 20 | 20 |
| *. Correlation is significant at the 0.05 level (2-tailed). | | | | |

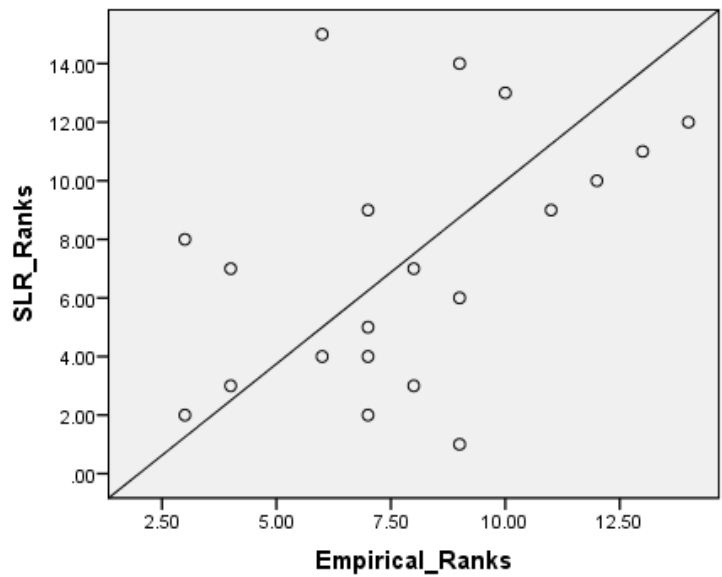

Figure 4.4 Scatter plot of the barriers ranks obtained from SLR and the survey questionnaire

## 4.4 Practices to address the identified critical success factors (CSFs) and critical barriers (CBs)

In order to address RQ7, I have used SLR to explore different practices in order to address the identified CSFs and CBs.

### 4.4.1 Practices identified using SLR

Using SLR, a total of 51 practices were identified to address the CSFs and CBs. In these practices, 27 practices were suggested for CSFs and 24 for CBs.

### 4.4.1.1 Practices identified using SLR for CSFs

The identified practices along the CFs are shown in Table 4.13. a total of 27 practices were identified from the primary studies to address the six most critical success factors.

In Table 4.13 total of five practices are discussed for 'CSF1: Users' and top management commitment'. In the identified practices 'P1-CSF1' is the most common practice and is cited by 58% of the selected studies.

Similarly, other practices were identified for critical success factors.

Table 4.13 Identified practice s for CSFs

| Practice No | CSF1: User's and Top Management Commitment |
|---|---|
| P1-CSF1 | Good management of resources and expenses to achieve a higher quality of the product. |
| P2-CSF1 | The manger should handle the organizational politics that could affect project development. |
| P3-CSF1 | The manager should assess and hire a trained developer for the development of HIS. |
| P4-CSF1 | Continuously monitor the development activities and validate the patient's requirement for HIS on each phase. |
| P5-CSF1 | Provide an interdisciplinary communication environment so that each team member should be aware of other related activities. |
| | **CSF2: 3C's (Communication, Coordination, Control)** |
| P1-CSF2 | Leaders should provide a platform where team members could communicate face-to-face or by video conferencing in remote areas. |
| P2-CSF2 | Organizing frequent meetings of team members to coordinate in the project effectively even at the temporal ground. |
| P3-CSF2 | A leader should teach the ways of effective communication, coordination among team members, and controlling the entire project. Ss |
| P4-CSF2 | Providing a platform where top-level team members could communicate with patients. |
| P5-CSF2 | Organizations should appoint liaison among the geographically distributed teams. |
| | **CSF3: Staff Training and Involvement** |
| P1-CSF3 | Engaging socially powerful players early. Physician, nursing, and executive engagement are crucial to evaluation success. |
| P2-CSF3 | Ensure evaluation continuity: Have a plan for employee turnover at both the participant and evaluation administration team levels. |
| P3-CSF3 | Consider organizational, team, or other factors that may help (or hinder) the effects of training (and thus evaluation outcomes) |
| P4-CSF3 | Environmental signals before, during, and after training must indicate that the trained KSAs (knowledge, skills, and attitudes) and the evaluation itself should be valued by the organization. |
| P5-CSF3 | A platform should be established by the Organization for the team which allows the team member to exchange and share novelty. |
| | **CSF4: Relationship between Team Members** |
| P1-CSF4 | Prevent misunderstandings and errors, leaders or managers should practice and teach effective communication among team members. |
| P2-CSF4 | Healthcare leaders should build mutual respect and trust mechanism with each interaction they experience with clients and staff members also for themselves. |

| P3-CSF4 | Eliminating the cultural, temporal, and linguistics barriers for effective communication and relationship among team members. |
|---|---|
| P4-CSF4 | Leaders should guide solidarity, self-introduction among team members, and resolving conflicts. |
| | **CSF5: Protection of Patient Privacy and Data Security** |
| P1-CSF5 | The developed system should ensure confidentiality, Integrity, and Availability or patient record |
| P2-CSF5 | Early detection of security risks and predicting threat source should be a higher priority |
| P3-CSF5 | A robust privacy-preserving algorithm should be employed that protect the patient's identity and location. |
| P4-CSF5 | Ensure the use of cryptographic algorithms in the medical data sharing device to the cloud. |
| | **CSF6: Software Process Improvement protocols and Procedures** |
| P1-CSF6 | Process improvement team members have a deep understanding of software quality and software processes. |
| P2-CSF6 | Expertise needs continuous improvement using training. |
| P3-CSF6 | The team consists of individuals having previous process improvement experience, knowledge, and necessary skills. |
| P4-CSF6 | Use the previous experience of process improvement projects in global software development. |

### 4.4.1.2 Practices identified using SLR for CBs

A total of 24 best practices were identified from the literature in order to address the six most critical barriers that could negatively influence the HIS development process. CB1: Distributed Healthcare Team management issue is the most critical factor and P1-CB1 is the most cited practice in the literature for CB1.

Table 4.13 Identified practice s for CBs

| Practice No | **CB1: Distributed Healthcare Team Management Issues** |
|---|---|
| **P1-CB1** | Develop a platform for team members to periodically check their intention and work towards the development of HIS at the distributed level. |
| **P2- CB1** | Management should provide training and resources for the effective development of HIS. |
| **P3- CB1** | Develop an organizational culture supporting healthcare teams that resolve linguistic, cultural barriers |
| **P4-CB1** | Leaders should recognize the areas of overlap and gaps in leadership roles at the distributed level and should provide clarity about role boundaries to avoid ambiguity. |
| **P5-CB1** | The organizational top manager should examine critically the allocation of resources to leadership activities at distributed sites. |
| **P6-CB1** | Use procedures that encourage information sharing among the whole team, such as checklists, briefings, and IT solutions. |
| **P7-CB1** | Ensure that every member should participate in the development according to their best suitable time. |
| | **CB2: Lake of Patient Engagement** |
| **P1-CB2** | A platform should be provided at each site that engages relevant patients while developing the healthcare information system. |
| **P2-CB2** | Collect survey data from the engaged patients from all sites based on the blueprint of the proposed healthcare information system. |
| **P3- CB2** | Arrange training and workshops to understand both the culture of the people participating in the development of the system. |

| | CB3: National Policies towards HIS |
|---|---|
| P1-CB3 | Ministry of health should be involved before the development of actual HIS. A feasibility report should be approved by the ministry of health. |
| P2-CB3 | Ensuring that change in government policies may not affect the development of HIS |
| P3-CB3 | The government should provide enough budget and human resources for HIS development. |
| | **CB4: Scope change and creep** |
| P1-CB4 | Leaders should involve the team in setting the scope of the project to clear what should they develop. |
| P2-CB4 | Collaborate closely with your team members and patients. Show them the work in progress, iterate, and proactively involve them throughout the journey. |
| P3-CB4 | Raise issues right away, when they happen (just make sure you have had time to work through some solutions first). |
| P4-CB4 | Project managers should include a change management process in the Scope Management plan. |
| P5-CB4 | Business analysts can contribute to clear scope with effective requirements elicitation and by analyzing and documenting clear, complete, and concise requirements. |
| | **CB5: Lake of Technological Resource** |
| P1-CB5 | The manager should plan and allocate all the required modern technological resources for HIS development. |
| P2-CB5 | The manger should manage the cost and time for the development of HIS |
| P3-CB5 | The strategy should be developed so that HIS development activities will not disturb the everyday schedule of the team members. |
| | **CB6: Uncertainties and rapidly changing technology** |
| P1-CB6 | Management should focus on the developed modern technology for their project |
| P2-CB6 | Implementation of modern technology and kicking back of old should be done through procedure and protocols |
| P3-CB6 | Continuously monitoring the effectiveness of modern implemented technology |

### 4.4.2 Practices identified using a survey questionnaire

To reply to RQ8, an online survey of HIS experts was conducted based on the procedures defined using the SLR method. The ranking of the established activities for CSFs and CBs using the survey methodology is shown in Table 4.15 and Table 4.17, respectively. The tables are divided into three major categories: 'Positive' (SA), 'Agree(A)), 'Negative' (Strongly Disagree (SD), 'Disagree (SD)), and 'Neutral' (NS). The positive column represents the percentage of respondents who agree with the behaviors found using the SLR methodology. The negative column represents the percentage of respondents who thought that the identified activities could not be successful in resolving the identified CSFs and CBs. In the neutral group, I listed the percentage of respondents who were not aware of the rating of the identified activities.

### 4.4.2.1 Practices identified using survey questionnaire for CSFs

The results presented in Table 4.15 demonstrate that most of the experts were positively agreed with the practices identified using the SLR approach. This is obvious from the 'Positive' category of Table 4.15 where most of the values are above 60% except for a few practices.

Table 4.15 Identified practice s for CSFs using a survey questionnaire

| S. No | (Expert Opinion=77) | | | | | | | |
|---|---|---|---|---|---|---|---|---|
| | +Ve | | | -Ve | | | Neutral | |
| | SA | A | % | SD | D | % | NU | % |
| P1-CSF1 | 27 | 35 | 81 | 4 | 6 | 13 | 5 | 6 |
| P2-CSF1 | 25 | 30 | 71 | 6 | 8 | 18 | 8 | 10 |
| P3-CSF1 | 29 | 36 | 84 | 5 | 5 | 13 | 2 | 3 |
| P4-CSF1 | 25 | 31 | 73 | 4 | 3 | 9 | 14 | 18 |
| P5-CSF1 | 28 | 27 | 71 | 8 | 8 | 21 | 6 | 8 |
| P1-CSF2 | 24 | 34 | 75 | 6 | 7 | 17 | 6 | 8 |
| P2-CSF2 | 28 | 38 | 86 | 5 | 3 | 10 | 3 | 4 |
| P3-CSF2 | 28 | 40 | 88 | 3 | 3 | 8 | 3 | 4 |
| P4-CSF2 | 22 | 32 | 70 | 7 | 10 | 22 | 6 | 8 |
| P5-CSF2 | 22 | 26 | 62 | 7 | 9 | 21 | 13 | 17 |
| P1-CSF3 | 23 | 35 | 75 | 6 | 8 | 18 | 5 | 6 |
| P2-CSF3 | 28 | 30 | 75 | 6 | 7 | 17 | 6 | 8 |
| P3-CSF3 | 23 | 29 | 68 | 7 | 8 | 19 | 10 | 13 |
| P4-CSF3 | 24 | 28 | 68 | 8 | 8 | 21 | 9 | 12 |
| P5-CSF3 | 24 | 34 | 75 | 5 | 8 | 17 | 6 | 8 |
| P1-CSF4 | 22 | 28 | 65 | 7 | 8 | 19 | 12 | 16 |
| P2-CSF4 | 20 | 28 | 62 | 6 | 10 | 21 | 13 | 17 |
| P3-CSF4 | 23 | 33 | 73 | 6 | 10 | 21 | 5 | 6 |
| P4-CSF4 | 22 | 32 | 70 | 7 | 11 | 23 | 5 | 6 |
| P1-CSF5 | 25 | 34 | 77 | 4 | 8 | 16 | 6 | 8 |
| P2-CSF5 | 28 | 38 | 86 | 3 | 3 | 8 | 5 | 6 |
| P3-CSF5 | 24 | 28 | 68 | 7 | 9 | 21 | 9 | 12 |
| P4-CSF5 | 22 | 35 | 74 | 5 | 6 | 14 | 9 | 12 |
| P1-CSF6 | 24 | 34 | 75 | 6 | 5 | 14 | 8 | 10 |
| P2-CSF6 | 22 | 30 | 68 | 8 | 10 | 23 | 7 | 9 |
| P3-CSF6 | 24 | 32 | 73 | 6 | 11 | 22 | 4 | 5 |
| P4-CSF6 | 26 | 38 | 83 | 3 | 6 | 12 | 4 | 5 |

### 4.4.2.2 Practices identified using survey questionnaire for CBs

The results presented in Table 4.16 demonstrate that most of the experts were positively agreed with the practices identified using the SLR approach. This is obvious from the 'Positive' category of Table 4.16 where most of the values are above 60% except for a few practices.

Table 4.17 Identified practice s for CBs using a survey questionnaire

| S. No | (Expert Opinion=77) | | | | | | | |
|---|---|---|---|---|---|---|---|---|
| | +Ve | | | -Ve | | | Neutral | |
| | SA | A | % | SD | D | % | NU | % |
| P1-CB1 | 25 | 37 | 81 | 2 | 9 | 14 | 4 | 5 |
| P2- CB1 | 23 | 32 | 71 | 8 | 5 | 17 | 9 | 12 |
| P3- CB1 | 27 | 38 | 84 | 6 | 3 | 12 | 3 | 4 |
| P4-CB1 | 22 | 34 | 73 | 5 | 3 | 10 | 13 | 17 |
| P5-CB1 | 26 | 29 | 71 | 9 | 6 | 19 | 7 | 9 |
| P6-CB1 | 26 | 32 | 75 | 5 | 7 | 16 | 7 | 9 |
| P7-CB1 | 25 | 41 | 86 | 4 | 3 | 9 | 4 | 5 |
| P1-CB2 | 30 | 38 | 88 | 2 | 3 | 6 | 4 | 5 |
| P2-CB2 | 20 | 34 | 70 | 13 | 3 | 21 | 7 | 9 |
| P3- CB2 | 21 | 27 | 62 | 8 | 9 | 22 | 12 | 16 |
| P1-CB3 | 27 | 31 | 75 | 5 | 9 | 18 | 5 | 6 |
| P2-CB3 | 24 | 34 | 75 | 4 | 9 | 17 | 6 | 8 |
| P3-CB3 | 21 | 31 | 68 | 8 | 7 | 19 | 10 | 13 |
| P1-CB4 | 26 | 26 | 68 | 8 | 8 | 21 | 9 | 12 |
| P2-CB4 | 27 | 31 | 75 | 3 | 10 | 17 | 6 | 8 |
| P3-CB4 | 24 | 26 | 65 | 6 | 9 | 19 | 12 | 16 |
| P4-CB4 | 21 | 27 | 62 | 7 | 9 | 21 | 13 | 17 |
| P5-CB4 | 21 | 35 | 73 | 5 | 13 | 23 | 3 | 4 |
| P1-CB5 | 20 | 34 | 70 | 4 | 14 | 23 | 5 | 6 |
| P2-CB5 | 25 | 34 | 77 | 4 | 7 | 14 | 7 | 9 |
| P3-CB5 | 27 | 38 | 84 | 5 | 3 | 10 | 4 | 5 |
| P1-CB6 | 25 | 28 | 69 | 8 | 9 | 22 | 7 | 9 |
| P2-CB6 | 22 | 37 | 77 | 6 | 5 | 14 | 7 | 9 |
| P3-CB6 | 24 | 34 | 75 | 6 | 5 | 14 | 8 | 10 |

## 4.5 Additional analysis of identified success factors and barriers using Analytical Hierarchy Process (AHP)

An analytical hierarchy process (AHP) technique has been performed additionally on the barriers and the success factors in order to rank and to check the effectiveness.

### 4.5.1 Analytical Hierarchy Process (AHP)

AHP is a multi-criteria decision-making process used for prioritizing the influencing factor(s) of a specific project. Various studies have already been adopted this method to prioritize the factors of their interest project [34, 35]. The same approach is adopted in this study in order to prioritize the investigated barriers of the requirements engineering process in the context of GHIS. Table 1 shows the phases of AHP.

Table 4.18: List of the AHP phase

| Sr. No | Phases of AHP |
|---|---|
| Phase 1 | Classifying the goal, categories (factors), and their corresponding barriers (sub-factors) as shown in Figure 2. |
| Phase 2 | Develop the pair-wise comparison matrix of barriers based on the expert's opinions. |
| Phase 3 | Calculate the priority weight of each barrier category and sub-category by using pair-wise comparisons. |
| Phase 4 | Check the consistency of the judgments. |
| Phase 5 | Rank the barriers in their corresponding categories (local ranking of barriers). |
| Phase 6 | Determine the global weights of barriers (final rank of barriers). |
| Phase 7 | Prioritizing the barriers. |

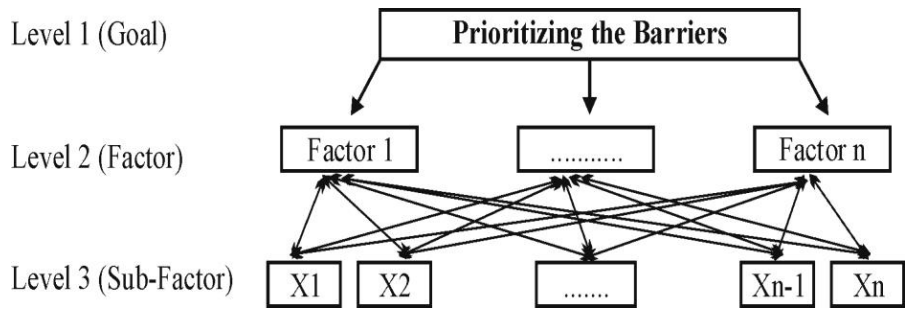

Figure 4.5: Hierarchy of the AHP

**Phase-1: (classifying the goal, categories (factors), and their corresponding barriers (sub-factors))**
In this phase, we decomposed a complicated decision-making problem into interlined decision elements [39, 40]. The complicated decision problem is divided at least three levels in the hierarchy structure. In this hierarchy structure, the basic goal of the decision-making problem is indicated at 1$^{st}$ level, the subcategories are indicated at level 2 and 3.

**Phase-2: (developed pair-wise comparison matrix of barriers on the basis of experts' opinion)**

At this level, we used a pair-wise comparison technique (Akbar et al 2020, Akbar et al 2017, Akbar et al 2017). The purpose of adopting the pair-wise technique is to calculate the weight (W) of each reported barrier and category of barriers. In a pair-wise approach, all barriers and their categories are compared with each other at every level based on their effect on the requirements engineering process in GHIS. On the basis of comparison, the pair-wise matrix was developed, and weight (W) was calculated. The weight (W) indicated the priority level of each barrier within the category and among the categories. The comparison matrix was developed (Akbar et al 2016, Akbar et al 2017, Akbar et al 2018) for each reported barrier and their categories as shown in matrix-A (n × n). The comparison matrix of the organization category is presented in Table 19.

Table 4.19: Description of the intensity scale

| Description | Significance intensity |
|---|---|
| Equally important | 1 |
| Moderately important | 3 |
| Strongly more important | 5 |
| Very strongly more important | 7 |
| Extremely more important | 9 |
| Intermediate values | 2, 4, 6, 8 |

$$\text{Matrix-A } (n \times n) \quad \begin{bmatrix} 1 & a12 & a1n \\ a21 & 1 & a2n \\ an1 & an2 & 1 \end{bmatrix}$$

**Phase-3 (calculation of priority weight (W) of each barrier)**

In order to determine the priority weights of each barrier, we adopted the following two steps:
- Normalization=Divide each element in every column by the sum of that column
- Weight (W)=Calculate the average of each row

**Phase-4 (determining the consistency of the matrix)**

**Analysis for Success factors:**

Table 4.20: Pairwise comparison for "Coordination" category

| S.No | SF1 | SF2 | SF3 | SF4 | SF5 | Weight(W) |
|---|---|---|---|---|---|---|
| SF1 | 1 | 2 | 1/3 | 2 | 1/2 | 0.146 |
| SF2 | 1/2 | 1 | 1/2 | 2 | 1/3 | 0.117 |
| SF3 | 3 | 2 | 1 | 7 | 4 | 0.438 |
| SF4 | 1/2 | 1/2 | 1/7 | 1 | 1/5 | 0.057 |
| SF5 | 2.00 | 3 | 0.25 | 5 | 1 | 0.242 |

λmax = 5.31, Consistency Index =0.08, Consistency Ratio (CR) = 0.07<0.1 (consistency OK)

Table 4.21: Pairwise comparison for "Human Resource Management" category

| S.No | SF6 | SF7 | SF8 | Weight(W) |
|---|---|---|---|---|
| SF6 | 1 | 2 | 1/2 | 0.312 |
| SF7 | 1/2 | 1 | 1/2 | 0.198 |
| SF8 | 2 | 2 | 1 | 0.490 |

λmax = 3.05, Consistency Index = 0.03, Consistency Ratio (CR) = 0.05<0.1 (consistency OK)

Table 4.22: Pairwise comparison for "Project Management" category

| S.No | SF9 | SF10 | SF11 | SF12 | SF13 | SF14 | Weight(W) |
|---|---|---|---|---|---|---|---|
| SF9 | 1 | 2 | 1/3 | 1/5 | 1/8 | 1/2 | 0.057 |
| SF10 | 1/2 | 1 | 1/2 | 1/5 | 1/7 | 1/2 | 0.046 |
| SF11 | 3 | 2 | 1 | 1/6 | 1/5 | 1/2 | 0.082 |
| SF12 | 5 | 5 | 6 | 1 | 2 | 9 | 0.426 |
| SF13 | 8 | 7 | 5 | 1/2 | 1 | 2 | 0.291 |

| SF14 | 2 | 2 | 2 | 1/9 | 1/2 | 1 | 0.099 |
|---|---|---|---|---|---|---|---|

λmax = 6.47, Consistency Index = 0.09, Consistency Ratio (CR) = 0.07 <0.1 (consistency OK)

Table 4.23: Pairwise comparison for "Technology" category

| S.No | SF15 | SF16 | SF17 | Weight(W) |
|---|---|---|---|---|
| SF15 | 1 | 1/2 | 8 | 0.413 |
| SF16 | 1/2 | 1 | 2 | 0.274 |
| SF17 | 0.125 | 1/2 | 1 | 0.158 |
| SF18 | 1/2 | 0.5 | 1/2 | 0.155 |

λmax = 4.27, Consistency Index = 0.09, Consistency Ratio (CR) = 0.09<0.1 (consistency OK)

Table 4.24: Pairwise comparison for "Categories"

| S.No | C1 | C2 | C3 | Weight(W) |
|---|---|---|---|---|
| C1 | 1 | 3 | 1/2 | 0.289 |
| C2 | 1/3 | 1 | 1/2 | 0.126 |
| C3 | 2 | 2 | 1 | 0.415 |
| C4 | 1/2 | 2 | 1/3 | 0.169 |

λmax = 4.17, Consistency Index = 0.06, Consistency Ratio (CR) = 0.06<0.1 (consistency OK)

**Phase-5: (Determining the local weight and global Weight)**

Table 4.25: Local and global ranking of success factor

| Categories | Weights of the categories | Success factors | Local weights | Local ranking | Global weights | Global ranking |
|---|---|---|---|---|---|---|
| Coordination | 0.289 | SF1 | 0.146 | 3 | 0.042 | 8 |
| | | SF2 | 0.117 | 4 | 0.034 | 12 |
| | | SF3 | 0.438 | 1 | 0.127 | 2 |
| | | SF4 | 0.057 | 5 | 0.016 | 18 |
| | | SF5 | 0.242 | 2 | 0.070 | 4 |
| Human Resource Management | 0.126 | SF6 | 0.312 | 2 | 0.039 | 10 |
| | | SF7 | 0.198 | 3 | 0.025 | 15 |
| | | SF8 | 0.49 | 1 | 0.062 | 6 |
| Project Management | 0.415 | SF9 | 0.057 | 5 | 0.024 | 16 |
| | | SF10 | 0.046 | 6 | 0.019 | 17 |
| | | SF11 | 0.082 | 4 | 0.034 | 11 |
| | | SF12 | 0.426 | 1 | 0.177 | 1 |
| | | SF13 | 0.291 | 2 | 0.121 | 3 |
| | | SF14 | 0.099 | 3 | 0.041 | 9 |
| Technology | 0.169 | SF15 | 0.413 | 1 | 0.070 | 5 |
| | | SF16 | 0.274 | 2 | 0.046 | 7 |
| | | SF17 | 0.158 | 3 | 0.027 | 13 |
| | | SF18 | 0.155 | 4 | 0.026 | 14 |

**Analysis of Barriers:**

Additional analysis for barriers has been performed on 17 barriers.

Table 4.26: Pair-wise comparison between the barriers of "organizational management" category

| S. No | B1  | B2   | B3   | B4   |
|-------|-----|------|------|------|
| B1    | 1   | 7    | 5    | 6    |
| B2    | 1/7 | 1    | 1/5  | 3    |
| B3    | 1/5 | 5    | 1    | 1/5  |
| B4    | 1/6 | 1/3  | 5    | 1    |
| Total | **1.51** | **13.33** | **11.20** | **10.20** |

Table 4.27: pair-wise comparison between the barriers of "human resources management" category

| S. No | B5  | B6  | B7  | B8  | B9  | Weights (W) |
|-------|-----|-----|-----|-----|-----|-------------|
| **B5** | 1   | 3   | 1/5 | 1/7 | 3   | 0.10        |
| **B6** | 1/3 | 1   | 1/3 | 1/5 | 3   | 0.09        |
| **B7** | 5   | 3   | 1   | 1/5 | 3   | 0.15        |
| **B8** | 7   | 5   | 5   | 1   | 1/2 | 0.14        |
| **B9** | 1/3 | 1/3 | 1/3 | 2   | 1   | 0.13        |

*Λ=5.37, CI=0.0925, RI=1.12, CR=0.08<0.1(Consistent)*

Table 4.28: pairwise comparison between the barriers of "coordination" category

| S. No | B10 | B11 | B12 | B13 | B14 | B15 | B16 | B17 | Weight (W) |
|-------|-----|-----|-----|-----|-----|-----|-----|-----|------------|
| **B10** | 1   | 1/3 | 4   | 3   | 1/5 | 2   | 1/7 | 1/3 | 0.04 |
| **B11** | 3   | 1   | 3   | 1/6 | 1/7 | 3   | 1/5 | 4   | 0.09 |
| **B12** | 1/4 | 1/3 | 1   | 3   | 1/5 | 1/3 | 1/4 | ½   | 0.03 |
| **B13** | 1/3 | 6   | 1/3 | 1   | 1/3 | 3   | 1/5 | 3   | 0.08 |
| **B14** | 5   | 7   | 5   | 3   | 1   | 7   | 2   | 1/3 | 0.17 |
| **B15** | 1/2 | 1/3 | 3   | 1/3 | 1/7 | 1   | 1/7 | 5   | 0.06 |
| **B16** | 7   | 5   | 4   | 5   | ½   | 7   | 1   | 5   | 0.11 |
| **B17** | 3   | 1/4 | 2   | 1/3 | 3   | 1/5 | 1/5 | 1   | 0.07 |

*Λ=8.65, CI=0.0928, RI=1.41, CR=0.067<0.1(Consistent)*

Table 4.29: Pair-wise comparison between the categories of the barriers

| Sr. No | Organizational management | Human resources management | Coordination | Weight (W) |
|--------|---------------------------|----------------------------|--------------|------------|
| **Organizational management** | 1   | 3   | 1/3 | 0.31 |
| **Human resources management** | 1/3 | 1   | 1/5 | 0.11 |
| **Coordination** | 2   | 5   | 1   | 0.58 |

*Λ=3.046, CI=0.023, RI=0.58, CR=0.0396<0.1(Consistent)*

Table 4.30: Local and global ranking of barriers

| Categories | Category Weights | Barriers | Local weights | Local ranking | Global weights | Global ranking |
|------------|------------------|----------|---------------|---------------|----------------|----------------|
| Organizational management | 0.30 | B1 | 0.35 | 1 | 0.111 | 1 |
|  |  | B2 | 0.10 | 6 | 0.033 | 8 |
|  |  | B3 | 0.9  | 7 | 0.030 | 9 |
|  |  | B4 | 0.10 | 6 | 0.033 | 8 |
| Human resources management | 0.10 | B5 | 0.09 | 7 | 0.010 | 15 |
|  |  | B6 | 0.08 | 8 | 0.008 | 16 |
|  |  | B7 | 0.14 | 3 | 0.015 | 12 |
|  |  | B8 | 0.13 | 4 | 0.014 | 13 |

|  |  | B9 | 0.12 | 5 | 0.013 | 14 |
|---|---|---|---|---|---|---|
|  |  | B10 | 0.02 | 12 | 0.022 | 10 |
|  |  | B11 | 0.08 | 8 | 0.051 | 4 |
|  |  | B12 | 0.02 | 13 | 0.018 | 11 |
| Coordination | 0.57 | B13 | 0.07 | 9 | 0.045 | 5 |
|  |  | B14 | 0.16 | 2 | 0.097 | 2 |
|  |  | B15 | 0.05 | 11 | 0.034 | 7 |
|  |  | B16 | 0.11 | 6 | 0.064 | 3 |
|  |  | B17 | 0.07 | 10 | 0.040 | 6 |

## 4.6 Summary

This chapter presents the findings and interpretation of the SLR and empiric research performed with HIS experts working in the context of GHIS. The data collected using the SLR and questionnaire surveys were analyzed to identify success factors that have a positive effect on HIS implementation and obstacles that have a negative impact on process improvement activities. I have evaluated the most critical success factors (CSFs) and critical barriers (CBs) using the identified criteria mentioned in section (4.2.4). Following these criteria, six performance factors have been described as the most important to the implementation of the HIS system in GHIS organizations. The identified CSFs are user's top management commitment, staff training, and involvement, 3C (communication, coordination, and control), the relationship between team members, the Protection of privacy and data security, software process improvement, and procedure.

Using the same criterion for CBs, a total of five barriers were listed as the most critical barriers. The selected CBs are: Distributed Healthcare Team, Lake of Technological Resources, Uncertainty and Rapid Technology Change, National HIS Strategy, Scope Change, and Patient Engagement Lake and are considered to be the most important barriers to HIS implementation.

I have also described different activities using the SLR and the answers obtained from the experts using the questionnaire survey. The activities recorded will be used to resolve the established CSFs and CBs. In addition, the defined success factors and obstacles were evaluated using the analytical hierarchy method (AHP).

# CHAPTER 5: Software Requirement Engineering Health-care Implementation Maturity Model for GHIS Organization

## 5.1 Introduction

This research study is conducted with the intention to develop the Software Requirement Engineering Health-care Implementation Maturity Model for the global health care information (GHIS) system that will help organizations for assessment and improvement of the Health care information system development program. SRE-HIMM is based on the SLR study that is conducted in the healthcare information system domain. Moreover, we have conducted Global software development (GSD) and an empirical study with HIS experts. This SRE-HIMM consists of three components such as SRE-HIMM maturity level components, SRE-HIMM factor component, and SRE-HIMM practices. In this chapter each SRE-HIMM component is briefly described and reported the results of three case studies conducted to evaluate the effectiveness of SRE-HIMM.

This chapter is structured as follows:

- In section 5.2, the structure of SRE-HIMM is discussed

- Section 5.3 reports the assessment dimension of SRE-HIMM

- Section 5.4 presents the assessment of SRE-HIMM.

- In section 5.5, the introduction of the organizations selected for case studies is reported.

- The analysis of the case studies is discussed in Section 5.6.

- In section 5.7, Summary results of case studies are reported.

- The feedback of the experts is presented in Section 5.8.

- Section 5.9 presents the summary of chapter

## 5.2 Structure of SRE-HIMM

This research study used two approaches such as SLR and survey questionnaire studies with the intention of identifying the success factors and the barriers that could have a positive and negative effect on the HIS implementation system in the context of GHIS.

I have implemented the concepts of available models that are CMMI (Chrissis, Konrad, & Shrum, 2006), IMM (Niazi, Wilson, &Zowghi, 2005) and SOVRM (Khan, 2011) to develop SRE-HIMM as shown in Figure 5.1, which illustrates the association of different components of SRE-HIMM. It shows how the findings of the SLR and empirical studies support for the development of the core three components of SRE-HIMM i.e.

- SRE-HIMM maturity level component
- SRE-HIMM factors (CSFs and CBs) component
- SRE-HIMM practices component

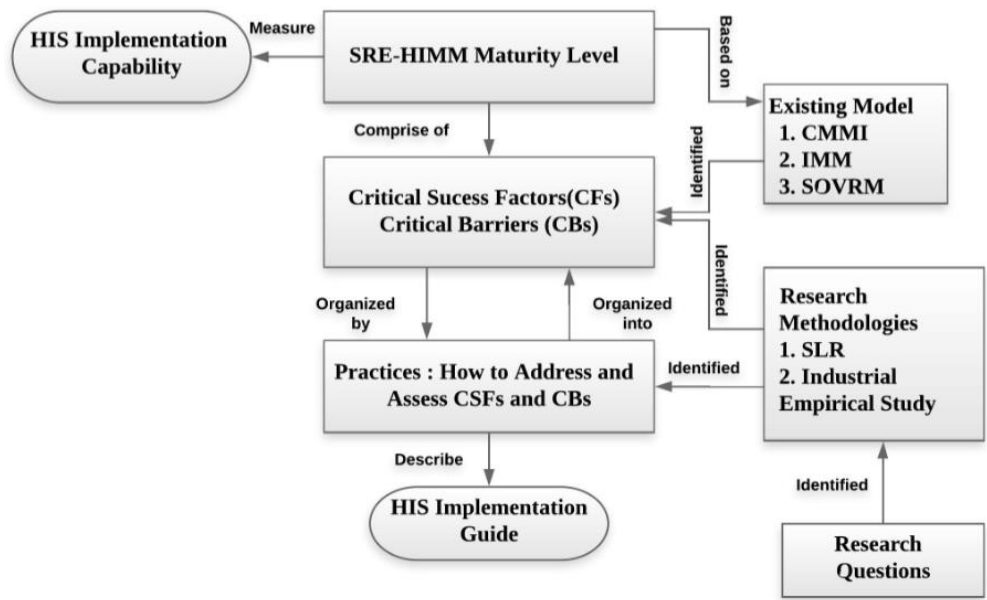

Figure 5.1: Structure of Proposed SRE-HIMM

These components are based on the existing models, i.e. CMMI (Chrissis, Konrad, & Shrum, 2003), IMM (Niazi, Wilson, &Zowghi, 2005) and SOVRM (Khan, 2011). CMMI has five maturity levels and every maturity level is based on different process areas (PAs). The process areas (PAs) of CMMI maturity levels contain various real-world practices. The CMMI maturity levels and its categories based on different PAs direct me to design SRE-HIMM maturity levels that are composed of different CSFs and CBs. The CSFs and CBs of maturity levels are identified using

SLR and survey questionnaire approaches. I have also identified the real-world practices with the intention to address each CSF and CB of a specific maturity level.

### 5.2.1 SRE-HIMM maturity level component

The staged representation of CMMI was observed in this research study in order to structure the SRE-HIMM maturity levels. For SREIMM, many changes have been made to the CMMI structure to accept the process improvement characteristics as shown in Figure 5.2:

- Level-1 (Initial): The level-1 of CMMI has been directly considered as the initial level of SRE-HIMM. In this level, the deployment process of HIS is chaotic, and few processes are defined. This level has no CSF and CB.
- Level-2 (Management Commitment): user's and top management has been reported as the most significant factor in the SLR study (75%) and survey questionnaire (88%). For the successful implementation of complex HIS, Top management should follow the long-standing approach and the commitment towards the development of HIS. The top management should provide adequate time and resources, manage internal politics, cultural, linguistic, and temporal issues. Thus, level-2 should be followed with the highest priority for the successful implementation of HIS. This level has one CSF and one CB as shown in Figure 5.3.
- Level-3 (Information Sharing): The 3'C (Communication, Coordination, Control) has been reported as the second most significant factor in the SLR (67%) study and survey questionnaire study (89%). At this level organizations should manage the issues that occurred due to the lake of communication among the management and the team members themselves. Strong coordination and control over the successful development of HIS comes with effective communication and sharing novelty work. Thus, this factor needs much intention to address at level-3 by organizations. Level-3 consists of two CSFs and one CB as described in Figure 5.3.
- Level-4 (Defined): This level has been adopted form the CMMI as in this level the organization should develop standards and procedures that will be implemented during the HIS development process. This level has two CSFs and two barriers as shown in Figure 5.3.
- Level-5 (Optimizing): Optimizing is the final maturity level of SRE-HIMM, this level is comprised of Level-4 and Level-5 of CMMI. At this level organization should involve the targeted patient for them the system is to be developed and continuously improve and respond to the change. This level consists of one CSF and one CB

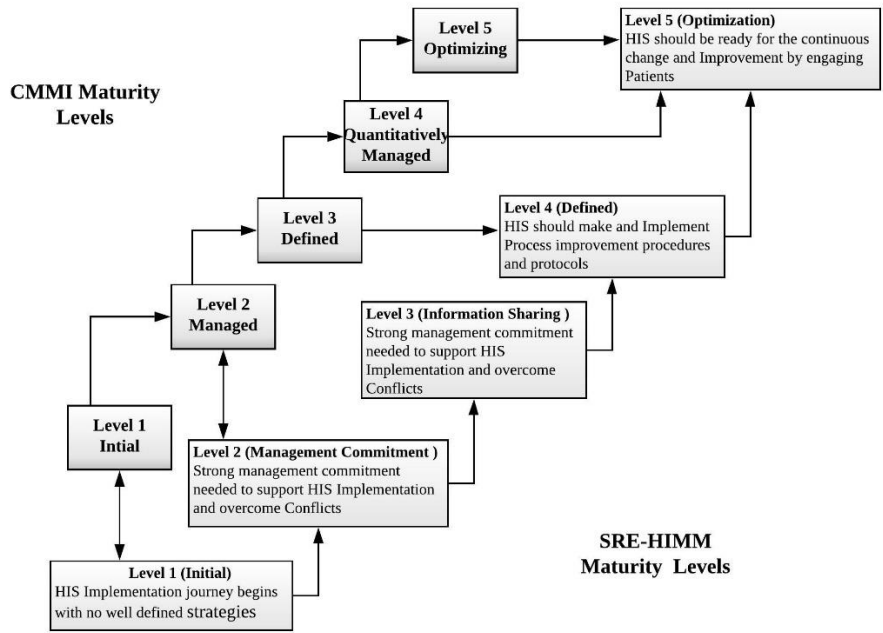

Figure 5.2: Proposed SRE-HIMM levels with CMMI Comparison

### 5.2.2 SRE-HIMM factors (CSFs and CBs) component

CMMI is comprised of 25 process areas (PAs) divided into five maturity stages. In IMM (Niazi, Wilson, & Zowghi, 2005) and SOVRM (Khan, 2011, Hamza et al. 2019, Mateen et al. 2018) the authors considered SPI 's maturity levels rather than process areas in terms of CSFs and CBs.

Several other researchers have followed the same idea by using the CSFs and CBs (Khan, 2015, Hayat et al. 2021, Qadri et al, 2022, Riaz et al 2022, Fahad et al 2021, Hamza et al 2022,). They used CSFs and CBs, instead of CMMI PAs. The importance of CSFs and CBs has been highlighted by numerous researchers (Niazi, Wilson, & Zowghi, 2005; Khan, 2011, Huotari, & Wilson, 2001; Khandelwal, & Ferguson, 1999; Khandelwal, & Natarajan, 2002; Pellow, & Wilson, 1993; Somers, & Nelson, 2001, Rafi et al. 2020, Shameem et al. 2020, Yasir et al 2020). Hence, the concept of using CSFs and CBs to develop SRE-HIMM was adopted.

Critical factors (CSFs and CBs) represent some of the important areas where organizations should focus during the implementation of HIS development process. The number of CSFs and CBs are identified using the criterion developed in (section 4.2.4). The identified CSFs and CBs are listed in Table 5.1.

Table 5.1 CSFs and CBs identified using SLR and survey questionnaires.

| S.No | Identified CSFs | S.No | Identified CBs |
|---|---|---|---|
| CSF 1 | User's and top Management Commitment | CB1 | Distributed Healthcare Team |

| CSF 2 | 3C's (Communication, Coordination, Control) | CB 2 | Lake of Patient Engagement |
| CSF 3 | Staff Training and Involvement | CB 3 | National Policies towards HIS |
| CSF 4 | Relationship between Team Members | CB 4 | Scope change and creep |
| CSF 5 | Protection of Patient Privacy and Data Security | CB 5 | Lake of Technological Resource |
| CSF 6 | Software Process Improvement protocols and Procedures | CB 6 | Uncertainties and rapidly changing technology |

CMMI process areas may be grouped into the categories of project management, engineering, service management, and support (Chrissis, Konrad, & Shrum 2003, Tahir et al 2020, Muhammad et al 2020). In this research report, the same methodology was adopted, and CSFs and CBs were classified. This classification of CSFs and CBs led me to design five maturity levels as shown in Figure 5.3, i.e. 'Initial,' 'Management Engagement,' 'Information Sharing,' 'Defined' and 'Optimization.' CSF and CB categorization was based on the current models in various other domains (Niazi, Wilson, & Zowghi, 2005; Khan, 2011; Khan, 2015; Ali and Khan, 2016, Kamal et al 2020, Shafiq et al 2018).

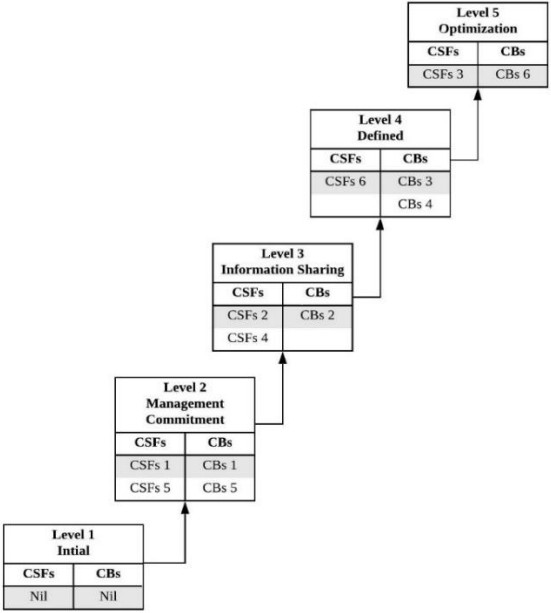

Figure 5.3 SRE-HIMM Maturity Levels corresponding to CSFs and CBs

### 5.2.3 SRE-HIMM practices component

To address the identified CSFs and CBs, I have identified a list of practices using SLR and a survey questionnaire as discussed in (section 4.4.1) and (section 4.4.2). All the identified practices along their respective CSFs are discussed in Table 5.2. Table 5.2 presents SRE-HIMM maturity levels with their respective CSFs and practices identified to address these CSFs. In Table 5.2, 'P-CSF' represents (practice for critical success factor). All the practices are briefly discussed in chapter 4 (Table 4.13, Table 4.16). Table 4.13 consists of those practices which were identified during the SLR study.

Table 5.2 Identified practice s for CSFs

| Maturity Levels | CSFs | Practices |
|---|---|---|
| Level-1: Initial | Nil | Nil |
| Level-2: Management Commitment | CSF1: Users' and Top Management commitment | P1-CSF1 (Practice is given in Table 4.13)<br>P2-CSF1 (Practice is given in Table 4.13)<br>P3-CSF1 (Practice is given in Table 4.13)<br>P4-CSF1 (Practice is given in Table 4.13)<br>P5-CSF1 (Practice is given in Table 4.13) |
| | CSF5: Protection of Patient Privacy and Data Security | P1-CSF5 (Practice is given in Table 4.13)<br>P2-CSF5 (Practice is given in Table 4.13)<br>P3-CSF5 (Practice is given in Table 4.13)<br>P4-CSF5 (Practice is given in Table 4.13) |
| Level-3: Information Sharing | CSF2: 3C's (communication, coordination, control) | P1-CSF2 (Practice is given in Table 4.13)<br>P2-CSF2 (Practice is given in Table 4.13)<br>P3-CSF2 (Practice is given in Table 4.13)<br>P4-CSF2 (Practice is given in Table 4.13)<br>P5-CSF2 (Practice is given in Table 4.13) |
| | CSF4: Relationship Between Team Members | P1-CSF4 (Practice is given in Table 4.13)<br>P2-CSF4 (Practice is given in Table 4.13)<br>P3-CSF4 (Practice is given in Table 4.13)<br>P4-CSF4 (Practice is given in Table 4.13) |
| Level-4: Defined | CSF6: Software Process Improvement protocols and Procedures | P1-CSF6 (Practice is given in Table 4.13)<br>P2-CSF6 (Practice is given in Table 4.13)<br>P3-CSF6 (Practice is given in Table 4.13)<br>P4-CSF6 (Practice is given in Table 4.13) |
| Level-5: Optimization | CSF3: Staff Training and involvement | P1-CSF3 (Practice is given in Table 4.13)<br>P2-CSF3 (Practice is given in Table 4.13)<br>P3-CSF3 (Practice is given in Table 4.13)<br>P4-CSF3 (Practice is given in Table 4.13)<br>P5-CSF3 (Practice is given in Table 4.13) |

Similarly, various practices were identified to address the identified CBs. The CBs along their respective practices are given in Table 5.3. In Table 5.3, 'P-CB' represents (practice for critical barrier).

Table 5.3 Identified practice s for CBs

| Maturity Levels | CBs | Practices |
|---|---|---|
| Level-1 (Initial) | Nil | Nil |
| Level-2 Commitment | CB1: Distributed Healthcare Team management issues | P1- CB1 (Practice is given in Table 4.14)<br>P2- CB1 (Practice is given in Table 4.14)<br>P3- CB1 (Practice is given in Table 4.14)<br>P4- CB1 (Practice is given in Table 4.14)<br>P5- CB1 (Practice is given in Table 4.14)<br>P6- CB1 (Practice is given in Table 4.14)<br>P7- CB1 (Practice is given in Table 4.14) |

|  | CB5: Lake of Technological Resources | P1-CB5 (Practice is given in Table 4.14)<br>P2-CB5 (Practice is given in Table 4.14)<br>P3-CB5 (Practice is given in Table 4.14) |
|---|---|---|
| Level-3 Information Sharing | CB2: Lack of Patient Engagement | P1- CB2 (Practice is given in Table 4.14)<br>P2- CB2 (Practice is given in Table 4.14)<br>P3- CB2 (Practice is given in Table 4.14) |
| Level-4: Defined | CB3: National Policies Toward HIS | P1-CB3 (Practice is given in Table 4.14)<br>P2-CB3 (Practice is given in Table 4.14)<br>P3-CB3 (Practice is given in Table 4.14) |
|  | CB4: Scope Change and Creep | P1-CB4 (Practice is given in Table 4.14)<br>P2-CB4 (Practice is given in Table 4.14)<br>P3-CB4 (Practice is given in Table 4.14)<br>P4-CB4 (Practice is given in Table 4.14)<br>P5-CB4 (Practice is given in Table 4.14) |
| Level-5 (Optimizing) | CB1: Uncertainties and rapidly changing technology | P1- CB1 (Practice is given in Table 4.14)<br>P2- CB1 (Practice is given in Table 4.14)<br>P3- CB1 (Practice is given in Table 4.14) |

## 5.3 SRE-HIMM assessment dimensions

To test SRE-HIMM as shown in Table 5.4, I have adopted the Motorola evaluation method (Daskalantonakis, 1994) Various other researchers have used the Motorola assessment tool to evaluate their proposed maturity models (Niazi, Wilson, & Zowghi, 2005; Khan, 2011; Khan, 2015; Ali, & Khan, 2016, Sardar et al. 2022, Nawaz et al, 2020, Shafiq et al 2020). The instrument for assessing Motorola was adopted for various compelling reasons. The instrument for evaluating Motorola is normative and has been tried and tested at Motorola. This instrument is used to determine an organization's current status compared to CMM and CMMI. This can show the weak areas of an organization that need to be addressed and improved (Daskalantonakis,1994). This is composed of the following three main dimensions of measurement (Daskalantonakis, 1994).

- **Approach:** The criterion of this dimension is the organizational management 's support and commitment to practice as well as an organization's ability to implement the practice.
- **Deployment:** This dimension's criterion is the clear and systematic application of practice in all areas of the project.
- **Results:** The criteria in this aspect is about the scope and consistency of positive outcomes over time and across the areas of the project.

A score was assigned from 0-10 for each dimension (Daskalantonakis,1994). For every dimension the scoring criterion is discussed in Table 5.4.

Table 5.4 Motorola assessment Instrument (Daska lantonakis, 1994)

| Score | Key Activity evaluation dimensions | | |
|---|---|---|---|
|  | **Approach** (Score Range: 0, 2, 4, 6, 8, 10) | **Deployment** (Score Range: 0, 2, 4, 6, 8, 10) | **Results** (Score Range: 0, 2, 4, 6, 8, 10) |
| Poor (0) | ➢ No management recognition of the need | ➢ No part of the organization uses the practice | ➢ Ineffective |
|  | ➢ No organizational ability<br>➢ No organizational commitment<br>➢ Practice not evident | ➢ No part of the organization shows interest |  |
| Weak (2) | ➢ Management begins to recognize the need<br>➢ Support items for the practice start to be created<br>➢ A few parts of an organization can implement the practice | ➢ Fragmented use<br>➢ Inconsistent use<br>➢ Deployed in some parts of the organization<br>➢ Limited to monitoring/verification of use | ➢ Spotty results<br>➢ Inconsistent results<br>➢ Some evidence of effectiveness for some parts of the organization |
| Fair (4) | ➢ Wide but not complete commitment by management<br>➢ Road map for practice implementation-defined<br>➢ Several supporting items for the practice in place | ➢ Less fragmented use<br>➢ Some consistency in the use<br>➢ Deployed in some major parts of the organization<br>➢ Monitoring/verification of use for several parts of the organization | ➢ Consistent and positive results for several parts of the organization<br>➢ Inconsistent results for other parts of the organization |
| Marginally qualified (6) | ➢ Some management commitment: some management becomes proactive<br>➢ Practice implementation well underway across parts of the organization<br>➢ Supporting items in place | ➢ Deployed in some parts of the organization<br>➢ Mostly consistent use across many parts of the organization<br>➢ Monitoring/verification of use for many parts of the organization | ➢ Positive measurable results in most parts of the organization<br>➢ Consistently positive results over time across many parts of the organization |
| Qualified (8) | ➢ Total management commitment<br>➢ Majority of management is proactive<br>➢ Practice established as an integral part of the process<br>➢ Supporting items encourage and facilitate the use of practice | ➢ Deployed in almost all parts of the organization<br>➢ Consistent use across almost all parts of the organization<br>➢ Monitoring/verification of use for almost all parts of the organization | ➢ Positive measurable results in almost all parts of the organization<br>➢ Consistently positive results over time across almost all parts of the organization |
| Outstanding (10) | ➢ Management provides zealous leadership and commitment<br>➢ Organizational excellence in the | ➢ Pervasive and consistent deployed across all parts of the organization | ➢ Requirements exceeded<br>➢ Consistently world-class results<br>➢ Counsel sought by others |

| | practice recognized even outside the company | ➢ Consistent use over time across all parts of the organization<br>➢ Monitoring/verification for all parts of the organization | |
|---|---|---|---|

The following steps of the Motorola instrument have been adopted for the SPIIMM assessment component (Daskalantonakis, 1994).

- Step-1: The participant of the HIS implementation team should compute the three-dimensional score of the Motorola instrument for each practice of critical success factors (CSFs) and critical barriers (CBs).
- Step-2: The calculated three-dimensional scores of each practice are summed together and divided by three (3). The final calculated score is rounding the nearest whole number.
- Step-3: Step-2 should repeat for each practice of the identified CSFs and CBs. Add the scores of all the practices together to calculate the final score for specific CSF or CB.
- Step-4: Relating the assessment score to SRE-HIMM: a score of 7 or above for a specific CSF or CB will show that a particular CSF or CB has been effectively implemented (Daskalantonakis, 1994). If the score of any CSF or CB is lower than seven then the implementation of that particular CSF or CB considered weak (Daskalantonakis, 1994).
- Step-5: To achieve a specific maturity level of SRE-HIMM, it is vital to address all the CSFs and CBs of that particular maturity level. For example, if the organization wants to attain the maturity level-3 of SRE-HIMM it is a must to address all the CSFs and CBs of level 3 and its average score should be seven or above.

In Table 5.5 the evaluation example of SRE-HIMM is shown by following the above five steps of Motorola instrument.

Table 5.5: Assessment examples of SRE-HIMM factors by Motorola Assessment Instrument

| **CSF1: Users' and Top Management Commitment** | | | | | |
|---|---|---|---|---|---|
| | | **Key Activity evaluation dimensions** | | | Average Score (Average of the dimension values) |
| **S. No** | **Practices** | Approach (0,2,4,6,8,10) | Deployment (0,2,4,6,8,10) | Results (0,2,4,6,8,10) | |
| P1-CSF1 | Strong Management of resources and cost to achieve higher Project quality. | 6 | 6 | 8 | 7 |
| P2-CSF1 | The manger should handle the organizational politics that could affect project development. | 4 | 2 | 2 | 3 |
| P3-CSF1 | The manager should assess and hire a trained developer for the development of HIS. | 6 | 4 | 4 | 5 |

| P4-CSF1 | Continuously monitor the development activities and validate the patient's requirement for HIS on each phase. | 4 | 4 | 2 | 3 |
|---|---|---|---|---|---|
| P5-CSF1 | Provide an interdisciplinary communication environment so that each team member should be aware of other related activities. | 6 | 8 | 8 | 7 |
| Total of 'average scores'= (add all the average score values at the rightmost column) | | | | | 25 |
| Overall score= Total of 'average score'/ total number of practices= (25/5=5) | | | | | 5 |

## 5.4 Assessment of SRE-HIMM

The case study approach was followed to evaluate the proposed SRE-HIMM model, as discussed in section 5.3. To assess the efficiency of SRE-HIMM, three case studies were performed in GHIS organizations

### 5.4.1 Assessment criteria

The following criteria have been adopted in order to assess the effectiveness of SRE-HIMM.

- **Ease of use**: The objective of this criterion is to evaluate how easily experts can use and understand the implementation of SRE-HIMM.
- **User satisfaction**: The aim of this criterion is to analyze the user satisfaction level with the results of SRE-HIMM.
- **Structure of SRE-HIMM**: The objective of this criterion is to analyze different components of SRE-HIMM and also overview the distribution of critical success factors and critical barriers across the maturity levels of SRE-HIMM.

The criteria are based on existing literature and the same criteria have been adopted by different other researchers (Niazi, Wilson, & Zowghi, 2005; Ali, & S. U. Khan, 2016, Rafi et al 2020, Nasrullah et al, 2020, Sang et al. 2018). The above criteria are followed in order to examine those areas where SRE-HIMM needs further improvements. All the evaluation results of SRE-HIMM are used to modify the structure of SRE-HIMM and to use it for future study.

### 5.4.2 Limitations of the assessment

The results discussed in this chapter are based on the three case studies that GHIS organizations conduct. SRE-HIMM 's assessment of this small sample size may not be effective for all GHIS organizations. A single representative responded to each case study, and that individual's feedback may not present the population of the entire HIS team. I adopted the Motorola instrument (Daskalantonakis, 1994) to assess the CSFs and CBs identified alongside their practices. It may not be enough to use the Single Evaluation Instrument to obtain more accurate outcomes. I also conducted all the case studies using email correspondence that provided no face-to-face interaction with the respondent. Due to a lack of face-to-face communication, the documents provided for the case study might not have been fully understood.

### 5.4.3 SRE-HIMM assessment using case studies

The case study approach was used in this research work to test the SRE-HIMM. The case study approach is considered to be more suitable for evaluation and may provide adequate information on the experiences of the real-world software industry (Yin, 2013). Since SRE-HIMM is designed to be applied in the real-world software industry, the case study approach is considered to be more relevant and efficient for this research. It also helps to highlight those SRE-HIMM areas that need further improvements. The case study approach helps to check SRE-HIMM's practicality and usability.

separate case studies were conducted in GHIS organization located in India. I have visited their websites for further information and detail about the HIS efforts and activities The selected organization gives a rich description of their process improvement activities and they were agreed to issue the results of the case studies.

Each participant was contacted through email in order to provide complete information and understanding of the case study. I have sent the following three main documents to each participant.

- Assessment of SRE-HIMM critical success factors (CSFs) and critical barriers (CBs) using the Motorola assessment Instrument
- Consent form

Guideline document consists of a brief description of the SRE-HIMM including the introduction of the research project, the structure of SPIIMM, and the assessment tool adopted for SRE-HIMM evaluation.

The second document consists of the identified CSFs and CBs along with their practices and each participant have to assess those CSFs and CBs for their respective organizations. The assessment is based on the Motorola evaluation instrument discussed in section 5.3. The assessment of SRE-HIMM critical success factors (CSFs) and critical barriers (CBs) document is presented in Appendix H.

The consent form was provided in order to conduct the feedback session with each participant regarding the SRE-HIMM. The questionnaire, as presented in Appendix I was used to structure the feedback session. The questionnaire consists of three sections i.e. Section-A, Section-B, and Section-C. Section-A provides detailed demographic information about each participant. Section-B contains the evaluation of SRE-HIMM according to the assessment criterion discussed in section 5.4.1. In Section-C, the list of the identified practices for CSFs and CBs is provided to the participant for review and suggestions. Each questionnaire of the feedback section was qualitatively analyzed due to the qualitative nature of the questionnaire questions.

## 5.5 Introduction of the organizations selected for case studies

The selected three companies are tagged as Company A, B, and C. Due to privacy reasons the original names of the responding organizations are not disclosed.

### 5.5.1 Profile of Company A

Company A is an ISO 9001:2008 certified international company that provides software and system development and services related to information technology. Company A is a large organization highly is a highly specialized provider of Medical Automation Software and Consulting, situated in Kerala. Their flagship product Blooray MeDIC is being used by thousands of medical professionals successfully. The participant of this case study is a product process consultant working in the branch of company A located in Delhi, India.

The aim of company A is to provide consultancy and technical services in the following major areas.

- EMR Developer
- Medical automation software and consultancy

### 5.6 Analysis of the case studies

The following analyses have been performed on the data collected during the case studies.

- Measure the maturity level of an organization for process improvement based on the assessment document provided to the respondents.
- According to the assessment document of SRE-HIMM a score of 7 or above for particular CSF or CB denotes that a specific factor has been successfully implemented. If the overall score is less than 7 then the implementation status of the factor will be marked as weak.
- The data collected during the feedback session will be used to analyze the effectiveness of the SRE-HIMM.
- The analysis of the feedback session will clarify whether SRE-HIMM is clear, easy to use, and contributing to enhancing the process improvement in the domain of GHIS.
- The feedback session will analyze the ease of use and understanding of each practice design to address the CSFs and CBs.
- The results of the case study data will help to check the generalization and applicability of SRE-HIMM.

### 5.6.1 Results at Company A

The SRE-HIMM assessment document as shown in Appendix H was provided to each participant in the case study. The assessment document was based on the Motorola assessment instrument (Daskalantonakis, 1994). The case study participant used SRE-HIMM and assessed the maturity level of the organization. The case study results obtained from company A are shown in Table 5.6 and shown in Appendix H1. The results shown in Table 5.6 show that Company A is at Level 3 of SRE-HIMM. Company A presented both Level-2 and Level-3 CSFs and CBs because the average score for each CSF and CB is $\geq 7$.

In order to achieve a specific maturity level, it is necessary to address all CSFs and CBs of that maturity level by having an average score $\geq 7$. Therefore, in order to achieve maturity level 4,

It is necessary for A to strongly address the critical success factor. This indicates that Company A paid much less attention to carrying out the experimentally focused implementation of newly implemented standards and models.

In Level-3, it is shown that Software Process Improvement protocols and Procedures have a high score of 8.83 which clarifies that company A has a well-established system to provide adequate resources for each project.

Similarly, to achieve Level-5 of SPIIMM, it is vital for company A to achieve all the previous levels and address the critical success factor (CB5: Uncertainties and rapidly changing technology) which have an average score of 6. It highlighted that company A is lacking experienced individuals working in the domain of software process improvement.

Table 5.6 Assessment results for each CSF and CB in company A

| Maturity Levels | CSFs and CBs | Final Score | Status in Company A |
|---|---|---|---|
| Level-1 (Initial) | - | - | - |
| Level-2 Management Commitment | CSF1: Users' and Top Management commitment | 7.8 | Strong |
| | CSF5: Protection of Patient Privacy and Data Security | 7.5 | Strong |
| | CB1: Distributed Healthcare Team management issues | 8 | Strong |
| | CB5: Lake of Technological Resources | 7.3 | Strong |
| Level-3 Information Sharing | CSF2: 3C's (communication, coordination, control) | 7.2 | Strong |
| | CSF4: Relationship Between Team Members | 7.25 | Strong |
| | CB2: Lack of Patient Engagement | 8.3 | Strong |
| Level-4 Defined | CSF6: Software Process Improvement protocols and Procedures | 6 | Week |
| | CB3: National Policies Toward HIS | 6.66 | Week |
| | CB4: Scope Change and Creep | 7.4 | Strong |
| Level-5 (Optimizing) | CSF3: Staff Training and involvement | 7.8 | Strong |
| | CB5: Uncertainties and rapidly changing technology | 7.8 | Strong |

Summarizing, the picture from analysis discussed in Table 5.6, company A needs to implement the weakly addressed factors at Level-4 and Level-5 in order to achieve the higher maturity i.e. Level-5 (Optimizing).

## 5.9 Summary

This chapter presents the complete architecture of SRE-HIMM that could help HIS organizations assess their level of maturity. SRE-HIMM is based on a systematic literature review analysis and an empirical study with HIS experts. The SRE-HIMM consists of three key components, i.e. SRE-HIMM maturity level component, SRE-HIMM factors (CSFs and CBs) component, and SRE-HIMM practice component. The objective of the maturity level component is to provide the HIS organization with an assessment level in order to know its development implementation capabilities. The SRE-HIMM factors (CSFs and CBs) component is structured to summarize the factors that can have positive or negative effects on HIS implementation in the domain of GSD. The SRE-HIMM practices component provides guidelines for organizations to understand and address their strong and weak areas for HIS implementation in GSD.

I have addressed the findings and interpretation of the case studies performed to test the industrial application of SRE-HIMM. In this case study, SRE-HIMM was tested by HIS experts and the findings were surprisingly good. They assessed the HIS maturity level of their firms using SRE-HIMM. He found SRE-HIMM to be a reliable model that could help HIS organizations determine their level of maturity and recommend improvement practices.

SRE-HIMM is dynamic in nature and I predict that in future it will be further refined and evolve based on the industrial employment and trends of GSD organizations.

# CHAPTER 6: CONCLUSION

## 6.1 Introduction

This chapter encapsulates the results and conclusions relevant to the implementation of the HIS development process, which drives conclusions and possibilities for future work that could be carried out. This chapter is organized according to the following:

## 6.2 Addressing the Research Questions

In this research study, we have developed a software requirement engineering healthcare implementation maturity model (SRE-HIMM). The model was developed by addressing a total of six questions. I have extracted the primary studies for the peer literature review that focus on our research interest. The primary studies were further analyzed, and I identified the success factors and barriers with their best practices. The identified success factors, barriers, best practices were further analyzed with the HIS experts. Furthermore, we have analyzed CMMI model critically and developed SRE-HIMM based on the CMMI levels. The model was developed based on critical success factors and critical barriers. The developed model was further analyzed with a case study performed with the company.

## 6.3 Limitations of the Study

I have adopted an SLR and mixed-method (questionnaire survey, case study) approaches for the development of SRE-HIMM in this research study.

SLR is used to define the factors and barriers to success that may influence the HIS development process. We also used the SLR approach to extract various practices to address the identified success factors and barriers. To extract the factors data a selection of 53 studies was chosen. With the numerous research articles on HIS, some related research papers may have missed out on this study.

The survey questionnaire was used with the HIS practitioners to verify the factors taken from the literature. I have not specifically checked the survey respondent 's perspectives and expectations. This may show that what the practitioners responded concerning the success factors and barriers may not necessarily be important for HIS activities. Perhaps their perceptions aren't accurate. Likewise, the sample size of the survey questionnaire (n=77) may not be high enough to explain the production and validity of SRE-HIMM, due to the limited time and resources. However, these are adequate samples based on the various other existing studies (Khan, 2015; Ali, & Khan, 2016) to justify the validity and assessment of SRE-HIMM.

SRE-HIMM 's evaluation of the limited sample size (n=1) of the case study may not be appropriate for all GHIS organizations. A single representative responded to each case study, and that individual's input may not reflect the population of the entire HIS team. I used the Motorola instrument (Daskalantonakis, 1994) to analyze the CSFs and CBs found along with their activities. Using the single-assessment instrument cannot be sufficient to obtain more accurate performance.

# ACKNOWLEDGMENT

Foremost, I would like to express my sincere gratitude to my supervisor Hu Chunqiang and Professor Haibo Hu for the continuous support of my master's study and research, for his patience, motivation, enthusiasm, and immense knowledge. I could not have imagined having a better advisor for my master's study.

Besides my Supervisors, I would like to thank my mentor, Dr. Muhammad Azeem Akbar for his invaluable guidance, instructions, encouragement, and full support that led my research work to success.

Last but not the least, my deepest gratitude goes to my father, Muhammad Sharif, Mother Razia BiBi, and my brother Abdul Razzaq for their endless love, patience, prayers, and encouragement throughout my career.

# Appendix:

All the appendixes are saved on this URL link in order to reduce the space

https://tinyurl.com/y7tylo2c

# Published papers:

Hamza, Muhammad, et al. "Identification and Prioritizing the barriers of Requirements Engineering Process in the domain of Global Health-Care Information System" Journal of Medical Imaging and Health Informatics
Special issue on Innovations in Large Scale Healthcare Application Development

Hamza, Muhammad, et al. "SIOT-RIMM: Towards Secure IOT-Requirement Implementation Maturity Model." Proceedings of the Evaluation and Assessment in Software Engineering. 2020. 463-468.